\documentclass[a4paper,11pt]{article}
\usepackage{pos}

\usepackage{slashed}
\usepackage{hyperref}
\usepackage{amsthm}
\usepackage{mathtools}
\usepackage{graphicx}
\usepackage{amsmath}
\usepackage{amsfonts}
\usepackage{simplewick}
\usepackage{tikz-cd}

\newcommand{\id}{\mathrm{id}}
\newcommand{\CF}{\mathcal{F}}

\newcommand{\ii}{{\mathrm{i}}}
\newcommand{\frv}{\mathfrak{v}}
\newcommand{\sfR}{\mathrm{R}}

\newcommand{\nn}{\nonumber}
\def\d{{\rm d}}

\def\RR{{\mathcal R}}

\newcommand{\sfh}{{\sf h}}
\newcommand{\sfp}{{\sf p}}
\newcommand{\sgreen}{\mathsf{G}}
\newcommand{\sP}{\mathsf{P}}
\newcommand{\sH}{\mathsf{H}}
\newcommand{\sI}{\mathsf{I}}

\def\BV{{\textrm{\tiny BV}}}
\newcommand{\Sym}{\mathrm{Sym}}

\newcommand{\BVL}{{\Delta}_{\textrm{\tiny BV}}}
\newcommand{\CS}{\mathcal{S}}
\newcommand{\tte}{\mathtt{e}}
\newcommand{\e}{{\mathrm{e}}}
\newcommand{\FR}{\mathbb{R}} 
\newcommand{\dd}{\mathrm{d}}

\newcommand{\iso}{\mathfrak{iso}}

\title {BV quantization of braided scalar field theory}

\author[a]{Dj. Bogdanovi\'c}
\author*[a]{M. Dimitrijevi\' c \'Ciri\'c}
\author[a]{V. Radovanovi\' c}
\author[b]{R. J. Szabo}

\affiliation[a]{University of Belgrade, Faculty of Physics\\
Studentski trg 12, Belgrade, Serbia}

\affiliation[b]{Department of Mathematics, Heriot--Watt University, Edinburgh, United
Kingdom\\
Maxwell Institute for
Mathematical Sciences, Edinburgh, United Kingdom\\
Higgs Centre
for Theoretical Physics, Edinburgh, United Kingdom}

\emailAdd{djbogdan@ipb.ac.rs}
\emailAdd{dmarija@ipb.ac.rs}
\emailAdd{rvoja@ipb.ac.rs}
\emailAdd{R.J.Szabo@hw.ac.uk}

\abstract{We address the problem of UV/IR mixing in noncommutative
quantum field theories from the perspective of braided $L_\infty$-structures and the Batalin--Vilkovisky formalism. We describe the example of
braided noncommutative scalar
field theory and its quantization using braided homological perturbation theory. The formalism is illustrated through one-loop calculations of the
two-point functions for $\phi^4$-theory in four dimensions and $\phi^3$-theory in six dimensions. In both cases we find that there are no
non-planar diagrams and no UV/IR mixing.}

\notes{\note{Preprint:  {\sf EMPG--23--07}}}

\FullConference{%
  Corfu Summer Institute 2022 "School and Workshops on Elementary Particle Physics and Gravity",\\
  28 August -- 1 October 2022\\
  Corfu, Greece
}

\tableofcontents

\begin{document}
\maketitle

\section{Introduction}

The quantization of noncommutative field theories is a long-standing problem. In Moyal--Weyl and other types of noncommutative field theories, non-planar Feynman diagrams appear in perturbation theory. At one-loop these diagrams are UV finite unless the external momentum is very small; in that case the UV divergences of the undeformed theories reappear. This behaviour is called UV/IR mixing and it is detrimental to the renormalizability of the field theory~\cite{Minwalla:1999px}, see e.g.~\cite{U1Reviewa} for a review. 

Intensive investigations over the years have been devoted to understanding this problem, see e.g.~\cite{Blaschke} for a review. To account for a (twist) deformed Poincar\'e symmetry, braided deformations of the oscillator algebras of field creation and annihilation operators were suggested in several works, see e.g.~\cite{Bal, Bu,Wess, Aschieri:2007sq}. Based on this, it was argued in \cite{Bal} that  noncommutative scalar quantum field theories with braided symmetry do not show UV/IR mixing in S-matrix elements. This lies in agreement with results of Oeckl's `braided quantum field theory'~\cite{Oeckl}, which however has not been extended to theories with gauge symmetries.

 Using the twist formalism, braided $L_\infty$-algebras and the associated braided noncommutative field theories have been constructed in~\cite{BraidedLinf}. The correspondence between $L_\infty$-algebras and the Batalin--Vilkovisky (BV) formalism, reviewed in e.g.~\cite{BVChristian}, was generalized to braided $L_\infty$-algebras in~\cite{SzaboAlex}, giving another route to braided quantum field theory that is now capable of treating gauge theories. The BV quantization of braided fuzzy scalar field theories was discussed in \cite{SzaboAlex}, while the BV quantization of braided scalar field theory as well as braided quantum electrodynamics was discussed in~\cite{UsQED}. 
 
 In this contribution we address the problem of quantization of noncommutative scalar field theory initiated in~\cite{UsQED} using the underlying braided $L_\infty$-algebra and its associated BV formalism; for simplicity, here we will work with the Moyal--Weyl deformation, but more general deformations can be studied analogously, see Appendix~\ref{app:Drinfeld}. In Section~\ref{sec:BVreview} we briefly review the connection between $L_\infty$-algebras and the BV formalism in field theory. In Section~\ref{sec:BQFTreview} we  introduce  braided noncommutative scalar field theory and quantize it using braided homological perturbation theory. We present two explicit examples in Section~\ref{sec:examples}: one-loop calculations of the two-point functions for braided $\phi^4$-theory on $\mathbb{R}^{1,3}$ and braided $\phi^3$-theory on $\mathbb{R}^{1,5}$. In both cases the two-point functions at one-loop are the same as in the commutative case, confirming the expectations of~\cite{Bal, Bu,Wess, Aschieri:2007sq, Oeckl}. This suggests that non-planar diagrams and UV/IR mixing may be absent altogether in braided scalar field theory.

\section{Homotopy algebras and BV quantization}
\label{sec:BVreview}

 In this section we briefly discuss $L_\infty$-algebras of field theory, their dual BV description and quantization via the procedure of ``homotopy transfer''. 

\paragraph{\bf $\boldsymbol{L_\infty}$-algebras.}

An $L_\infty$-algebra is a $\mathbb{Z}$-graded vector space {\small$V=\bigoplus_{k\in 
\mathbb{Z}}\, V_{k}$} with a sequence of graded antisymmetric multilinear maps of degree $2-n$ for $n\geq1$ called $n$-brackets:
\begin{align}
\ell_{n}:  V^{\otimes n} \longrightarrow & V \ , \quad  v_1\otimes \cdots\otimes v_n \longmapsto 
\ell_{n} 
(v_1,\dots,v_n) \nn\\[4pt]
\ell_{n} (\dots, v,v',\dots) &=\>  -(-1)^{|v|\,|v'|}\, \ell_{n} (\dots, v',v,\dots) \ ,\nn
\end{align}
 where $|v|$ denotes the degree of a homogeneous element $v\in V$. The $n$-brackets must also fulfil a sequence of relations called homotopy Jacobi identities. The first three relations are given by
\begin{align}
&\underline{n=1:}  \quad \ell_{1}\big(\ell_{1}(v)\big) = 0 \ ,\label{Cochain}\\[4pt]
&\underline{n=2:}  \quad \ell_{1}\big(\ell_{2}(v_1,v_2)\big) = \ell_{2}\big(\ell_{1}(v_1),v_2\big) + (-1)^{|v_1|}\, 
\ell_{2}\big(v_1, \ell_{1}(v_2)\big)\ , \label{Leibniz}\\[4pt]
&\underline{n=3:} \quad  \ell_{1}\big(\ell_{3}(v_1,v_2,v_3)\big) =  - \ell_{3}\big(\ell_{1}(v_1),v_2,v_3\big) - (-1)^{|v_1|}\, 
\ell_{3}\big(v_1, \ell_{1}(v_2), v_3\big) \nn\\
& \hspace{5cm}- (-1)^{|v_1|+|v_2|}\, \ell_{3}\big(v_1,v_2, \ell_{1}(v_3)\big) \nn\\
&\hspace{5cm}-\ell_{2}\big(\ell_{2}(v_1,v_2),v_3\big) -(-1)^{(|v_1|+|v_2|)\,|v_3|}\, \ell_{2}\big(\ell_{2}(v_3,v_1),v_2\big) \nn\\
&\hspace{5cm} -(-1)^{(|v_2|+|v_3|)\,|v_1|}\, \ell_{2}\big(\ell_{2}(v_2,v_3),v_1\big) \ . \label{eq:homotopyJacobi}
\end{align}

 Cyclic $L_\infty$-algebras contain an additional structure called a cyclic pairing that is a graded symmetric and non-degenerate bilinear map $\langle-,-\rangle:V\otimes V\to\mathbb{R}$ of degree $-3$, satisfying
\begin{align}
\langle a_0,\ell_n(a_1,a_2,\dots,a_n)\rangle = (-1)^{n+(|a_0|+|a_n|)\,n+|a_n| \,\sum_{i=0}^{n-1}\,|a_i|} \ \langle 
a_n,\ell_n(a_0,a_1,\dots,a_{n-1})\rangle \ , \label{Pairing}
\end{align}
for all $n\geq1$.

As discussed in \cite{HohmZwiebach}, the data of any classical field theory are completely encoded in a corresponding $L_\infty$-algebra. In this contribution we restrict to theories that have only irreducible gauge symmetries, for which $$V=V_0\oplus V_1\oplus V_2\oplus V_3\ ,$$ where $V_0$ contains gauge parameters, $V_1$ contains gauge fields, $V_2$ contains the equations of motion, and $V_2$ contains the second Noether identities. 

\paragraph{\bf $\boldsymbol{L_\infty}$-algebra of scalar field theory.}

We illustrate this formalism on the very simple example of a real massive scalar field $\phi$ in four dimensions with $\lambda\,\phi^4$-interaction. This theory has no gauge symmetries and therefore $V= V_{1}\oplus V_{2}$, where $V_1=V_2=C^\infty(\FR^{1,3})$. The  nonvanishing brackets are given by
\begin{align}
\ell_1(\phi) & = \big(-\Box - m^2\big)\,\phi \ , \label{eq:l1}\\[4pt]
\ell_3(\phi_1, \phi_2, \phi_3) &= \lambda\, \phi_1\, \phi_2\, \phi_3 \ , \label{eq:intbrackets}
\end{align}
for $\lambda\in\FR$ and $\phi,\phi_1,\phi_2,\phi_3\in V_1$. 
The cyclic pairing of degree $-3$ is taken to be
\begin{align}
\langle\phi, \phi ^+\rangle & = \int\d^4 x \ \phi \, \phi^+ \label{eq:scalarpairing}
\end{align}
where $\phi\in V_1$ and $\phi^+\in V_2$. 

This construction leads to the action functional
\begin{align}
\begin{split}
S(\phi) & = \frac{1}{2!} \, \langle \phi, \ell_1(\phi)\rangle 
-\frac{1}{3!} \, \langle\phi, \ell_2(\phi, \phi)\rangle
- \frac{1}{4!} \, \langle\phi, \ell_3(\phi, \phi, \phi)\rangle +\cdots\\[4pt]
& = \int\d^4 x \ \left(\frac{1}{2}\,\phi\,\big(-\Box - m^2\big)\,\phi - \frac{\lambda}{4!}\,\phi^4\right) \label{eq:S0scalar}
\end{split}
\end{align}
and the equation of motion
\begin{align*}
F_\phi & = \ell_1(\phi) - \frac{1}{2}\,\ell_2(\phi, \phi) -\frac{1}{3!}\,\ell_3(\phi, \phi, \phi)+\cdots  = \big(-\Box - m^2\big)\,\phi - \frac{\lambda}{3!}\,\phi^3 = 0 \ .
\end{align*}
This can be extended in the evident way to scalar field theories in any dimension with arbitrary polynomial interactions.

\paragraph{\bf BV formalism and $\boldsymbol{L_\infty}$-algebras.}

The quantization of field theories possessing gauge symmetries is not straightforward. The standard Faddeev--Popov quantization procedure works well for a large number of theories, such as quantum electrodynamics and Yang--Mills theories. It is based on the path integral approach where ghost fields appear after gauge-fixing. These are scalar fields with fermionic statistics and they are necessary to keep the gauge-fixed theory unitary. The gauge-fixed action still possesses a nilpotent odd global symmetry, the Becchi--Rouet--Stora--Tyutin (BRST) symmetry. Physical states lie in the cohomology of the generator of BRST symmetry, $Q_{\textrm{\tiny BRST}}$. 

Faddeev--Popov quantization was later generalized to the Batalin--Vilkovisky (BV) quantization method. One of the main advantages of the BV formalism is that it can be applied to more general field theories, such as theories with reducible and/or open gauge algebras. In particular, the BV formalism was used in quantization of open and closed bosonic string field theory as well as topological field theories.

BV quantization has manifest gauge invariance and introduces antifields as sources for this symmetry. An antifield is introduced for each field and ghost, doubling the number of degrees of freedom in this way. The antibracket $\{ -,- \}$ is an odd non-degenerate Poisson bracket on the space of fields and antifields. The original classical action $S$ is extended to a new BV action
$S_\BV$ that satisfies the classical master equation $\{ S_\BV , S_\BV\} =0$, which reproduces in a compact way the
gauge structure of the original theory. 

To quantize the theory, the classical master equation is modified to the quantum master equation 
\begin{equation}\nn
\frac{1}{2}\,\{ S_\BV , S_\BV\} -\ii\, \hbar\, \BVL S_\BV=0 \ , 
\end{equation}
where now $S_\BV$ represents the quantum BV action and a new operator $\BVL$ is introduced, called the BV Laplacian. To calculate correlation functions and scattering amplitudes, gauge-fixing has to be performed. In the BV formalism, gauge-fixing is implemented by choosing a Lagrangian submanifold of the space of fields and antifields; this eliminates antifields in terms of functionals of fields. When appropriately implemented, the usual Feynman diagrammatic methods can be used. For a detailed review of the BV formalism, see \cite{Gomis94}.

From an algebraic perspective, the BV formalism constructs a differential graded commutative algebra\footnote{The shifted vector space $V[1]$ has the same underlying vector space as $V$ but with the degrees of its homogeneous
subspaces shifted by 1. For example, an element $A\in V[1]_1$ has degree $1-1=0$.} 
$\big(\Sym(V[1])^*,Q_\BV\big)$ for a graded vector space $V=V_0\oplus V_1\oplus V_2\oplus V_3$ which encodes the BV fields (ghosts and fields associated to a given gauge field theory) together with their antifields. The action of the BV differential $Q_\BV$ on fields $A\in V_1$ and ghosts $c\in V_{0}$ encodes the kinematical gauge symmetry of the field theory, that is, gauge transformations and closure of the gauge algebra. The action of $Q_\BV$ on the antifields $A^+\in V_2$ incorporates the dynamical brackets of the $L_\infty$-algebra, while the BV transformations of the antifields $c^+\in V_{3}$ correspond to the second Noether identities and the corresponding actions of the gauge parameters. The cohomology of $Q_\BV$ in degree~$0$ thus simultaneously encodes the quotients of the space of fields by the equations of motion and by the action of gauge transformations. These are exactly the classical observables of the field theory. 

 This data is supplemented by the antibracket $\{-,-\}$, which is the canonical graded Poisson bracket compatible with the differential $Q_\BV$, whose inverse is the symplectic form $\omega_\BV$ of degree~$-1$ on $V$. Using this non-degenerate symplectic pairing we identify the dual $V^*\simeq V[1]$ and
\begin{align}\nn
\Sym(V[1])^*\simeq \Sym(V[2]) \ .
\end{align}
Finally, the algebra of classical observables is fully specified by $\big(\Sym(V[2]),Q_\BV, \{-,-\}\big)$.

 Let us now briefly describe how to extract the corresponding $L_\infty$-algebra, following~\cite{BVChristian}. The action of the BV differential is given by taking the antibracket with the BV action functional $S_\BV$, $Q_\BV=\{S_{\BV},-\}$. The action functional can be expanded as
\begin{align}\nn
S_\BV = \sum_{m\geq2} \, S_\BV^{(m)} \ ,
\end{align}
where $S_\BV^{(m)}$ is the part of $S_\BV$ which is a polynomial of degree~$m$ in the BV fields. Then the brackets $\ell_n$ of the $L_\infty$-algebra are given by $\big\{S_\BV^{(n+1)},-\big\}$ for $n\geq1$. Nilpotency $(Q_\BV)^2=0$, or equivalently the classical master equation $\{S_\BV,S_\BV\}=0$, then translates to the homotopy Jacobi identities for the brackets $\ell_n$, and $(V,\{\ell_n\})$ is an $L_\infty$-algebra. The $(-1)$-shifted symplectic structure $\omega_\BV$ induces a cyclic pairing of degree $-3$ on $V$, making it into a cyclic $L_\infty$-algebra. 

 Conversely, starting from an  $L_\infty$-algebra $(V,\{\ell_n\})$ with a cyclic structure $\langle-,-\rangle$ of degree~$-3$, choose a basis $\{\tte_k\}\subset V$ and the corresponding dual basis $\{\tte^k\}\subset V^*\simeq V[3]$ such that $\langle\tte^k,\tte_l\rangle = \delta^k_l$ for all $k,l$. Following~\cite{BVChristian}, one introduces the ``contracted coordinate functions'' as the elements\footnote{Throughout this contribution we assume the Einstein summation convention over repeated upper and lower indices.}
\begin{align}\label{eq:contractedcoord}
\xi := \tte^k\otimes \tte_k \ \in  \ \Sym(V[2])\otimes V
\end{align}
of degree~$1$. 

The $L_\infty$-structure on $V$ naturally extends to the tensor product $\Sym(V[2])\otimes V$ through the extended brackets
\begin{align}\nn
\ell_n^{\,\rm ext}(a_1\otimes v_1,\dots,a_n\otimes v_n) := \pm\,(a_1\odot\cdots\odot a_n)\otimes\ell_n(v_1,\dots,v_n) \ ,
\end{align}
for all $n\geq1$, $a_1,\dots,a_n\in\Sym(V[2])$ and $v_1,\dots,v_n\in V$; the explicit Koszul sign factors $\pm$ depend on the gradings, see \cite{BVChristian}. Similarly, the cyclic structure naturally extends to a symmetric non-degenerate pairing $\langle-,-\rangle^{\rm ext}:\big(\Sym(V[2])\otimes V\big)\otimes \big(\Sym(V[2])\otimes V\big)\to \Sym(V[2])$ of degree~$-3$ given by
\begin{align}\nn
\langle a_1\otimes v_1, a_2\otimes v_1\rangle^{\rm ext} := \pm\,(a_1\odot a_2)\,\langle v_1,v_2\rangle \ .
\end{align}

 With the contracted coordinate functions \eqref{eq:contractedcoord}, we now set
\begin{align*}
Q_\BV\xi = - \sum_{n\geq1} \, \frac{(-1)^\frac{n\,(n-1)}{2}}{n!} \, \ell_n^{\,\rm ext}(\xi^{\otimes n})  \ .
\end{align*}
The homotopy Jacobi identities imply that $(Q_\BV)^2=0$. The
extended pairing $\langle-,-\rangle^{\rm ext}$ induces a $(-1)$-shifted symplectic structure 
\begin{align*}
\omega_\BV := -\frac12\,\langle\delta\xi,\delta\xi\rangle^{\rm ext} \ \in \ \Omega^2(V[1]) 
\end{align*}
whose inverse defines the corresponding antibracket. In this way we recover the differential $Q_\BV$ and antibracket $\{-,-\}$ of the BV formalism directly from the cyclic $L_\infty$-algebra of a field theory. 

The BV action functional is obtained from the element 
\begin{align*}
S_\BV = \sum_{n\geq1} \, \frac{(-1)^\frac{n\,(n-1)}{2}}{(n+1)!} \, \langle\xi,\ell_n^{\,\rm ext}(\xi^{\otimes n})\rangle^{\rm ext} \ \in \ \Sym(V[2])
\end{align*}
of degree~$0$. Then
\begin{align*} 
Q_\BV = \{S_\BV,-\} \ ,
\end{align*}
and nilpotency $(Q_\BV)^2=0$ is equivalent to the {classical master equation}
\begin{align}\nn
\{S_\BV,S_\BV\} = 0 \ .
\end{align}

\paragraph{\bf Homological perturbation theory.}

Let us go back to the original $L_\infty$-algebra $(V,\{\ell_n\})$ describing the classical field theory. The propagators $\sfh$ of the free theory define 
a strong deformation retract, i.e.~a homotopy equivalence, of the cochain complex $(V,\ell_1)$ onto its cohomology
$H^\bullet(V)$:
\begin{equation}\label{SDR1}
\begin{tikzcd}
\arrow[out=120,in=60,loop,looseness=3,"\sfh"] (V,\ell_1) \ar[r,shift right=1ex,swap,"\sfp"] & \ar[l,shift right=1ex,swap,"\iota"] (H^\bullet(V),0) 
\end{tikzcd} \ .
\end{equation}
Here we introduced
\begin{itemize}
\item Inclusion: a cochain map $\iota :H^\bullet(V)\to V$ of degree~$0$;
\item Projection: a cochain map $\sfp : V\to H^\bullet(V)$ of degree~$0$;
\item Contracting homotopy $\sfh: V\to V$ of degree~$-1$.
\end{itemize}
These maps are required to satisfy the following conditions:
\begin{itemize}
\item $\sfp \, \iota = \id_{H^\bullet(V)}$;
\item $\iota\, \sfp - \id_{V} =  \ell_1\,\sfh + \sfh\,\ell_1$;
\item $\sfh^2=0$,  $\sfh\, \iota=0$ and $\sfp\, \sfh=0$.
\end{itemize}

We now extend this structure to the algebra of classical observables $\big(\Sym(V[2]),\ell_1, \{-,-\}\big)$ of the free theory. The antibracket is the graded Poisson bracket $$\{-,-\}:\Sym (V[2])\otimes \Sym (V[2])\longrightarrow \Sym (V[2])[1]$$ which is also a  graded derivation on $\Sym (V[2])$ in each of its slots; for example
\begin{align}\label{eq:der}
\{a_1,a_2\odot a_3\} = \langle a_1, a_2\rangle\odot a_3 \pm a_2\odot\langle a_1,a_3\rangle \ ,
\end{align}
for $a_1, a_2, a_3\in \Sym( V[2])$.
The antibracket is compatible with the differential $\ell_1$, extended as a graded derivation to all of $\Sym (V[2])$, as a consequence of cyclicity of the inner product $\langle-,-\rangle$. 

In this way we obtain a new strong deformation retract
\begin{equation}\label{SDR2}
\begin{tikzcd}
\arrow[out=160,in=20,loop,looseness=2,"\sH"] (\Sym(V[2]),\ell_1) \ar[r,shift right=1ex,swap,"\sP"] & \ar[l,shift right=1ex,swap,"\sI"] (\Sym (H^\bullet(V[2])),0) 
\end{tikzcd} \ .
\end{equation}
The maps $\sP$, $\sI$, and $\sH$ are  extensions of the corresponding maps in (\ref{SDR1}). We will present more details on how these extensions are realized in a particular example in Section~\ref{sub:BVscalar}. 

To incorporate interactions and quantize the theory, we use the homological perturbation lemma~\cite{Doubek:2017naz}. We perturb the differential $\ell_1$ to $Q_\BV = \ell_1 + \delta$, where $\delta$ is a small perturbation. This leads to a new strong deformation retract with new maps $\tilde{\sP}$, $\tilde{\sI}$, and $\tilde{\sH}$:
\begin{equation}\label{SDR3}
\begin{tikzcd}
\arrow[out=160,in=20,loop,looseness=2,"\tilde{\sH}"] (\Sym(V[2]),Q_\BV) \ar[r,shift right=1ex,swap,"\tilde{\sP}"] & \ar[l,shift right=1ex,swap,"\tilde{\sI}"] (\Sym (H^\bullet(V[2])),\tilde{\delta}) 
\end{tikzcd}\ .
\end{equation}
In particular, the new projection map is given by $\tilde{\sP}= \sP + \sP_{\delta}$ with
\begin{align*}
\sP_{\delta} = \sP\,\big(\id_{\Sym (V[2])} - {\delta}\,\sH\big)^{-1} \, \delta \, \sH \ .
\end{align*}
It was shown in \cite{Doubek:2017naz} that $\sP_{\delta}$ gives the path integral. 

Then the $n$-point correlation functions of the quantum field theory are defined as
\begin{align}\label{eq:BQFTnpoint}
\begin{split}
G_n(x_1,\dots,x_n) &= \langle0|{\rm T}[A(x_1)\cdots A(x_n)]|0\rangle := \sP_{\delta}(\delta_{x_1}\odot\cdots\odot\delta_{x_n}) \\[4pt]
&= \sum_{m=1}^\infty \, \sP\,\big(({\delta} \,\sH)^m\,(\delta_{x_1}\odot\cdots\odot\delta_{x_n})\big) \ ,
\end{split}
\end{align}
where $\delta_{x_a}(x):=\delta(x-x_a)$ are Dirac distributions supported at the insertion points $x_a$ of the physical field $A\in V_1$.  We will consider two important perturbations $\delta$. 

The free quantum field theory is defined by $\delta = \ii\,\hbar\,\BVL$ with the BV Laplacian $$\BVL:\Sym( V[2])\longrightarrow \Sym (V[2])[1] \ .$$
The BV Laplacian satisfies the two key properties $(\BVL)^2=0$ and $\BVL\circ\ell_1 = -\ell_1\circ\BVL$ which guarantee that $Q_\BV=\ell_1+\ii\,\hbar\,\BVL$ is a differential, $(Q_\BV)^2=0$. It is also related to the antibracket through
\begin{align}\label{eq:BVLantibracket}
\BVL(a_1\odot a_2) = \BVL(a_1)\odot a_2 + (-1)^{\vert a_1\vert}\, 
a_1\odot\BVL(a_2) +  \{a_1,a_2\} \ ,
\end{align}
for all $a_1,a_2\in \Sym (V[2])$.

The interacting quantum field theory is defined by $\delta = \ii\,\hbar\,\BVL + \{\CS _{\rm int},-\}$, where $\CS_{\rm int}$ is the interaction part of the BV action functional $S_\BV$. Then \eqref{eq:BQFTnpoint} is a formal power series in $\hbar$ and in the coupling constants appearing in $\CS_{\rm int}$, representing the perturbative expansion.

\section{Braided scalar quantum field theory}
\label{sec:BQFTreview}

In this section we apply BV quantization to braided noncommutative scalar field theory. For concreteness, here we will work with $\phi^4$-theory on ${\mathbb R}^{1,3}$. In Section~\ref{sub:braidedphi3}  we will additionally consider the example of $\phi^3$-theory in six dimensions.

\subsection{Braided $L_\infty$-algebra of scalar field theory}

Braided noncommutative field theories~\cite{BraidedLinf} are constructed using the twist formalism briefly reviewed in Appendix~\ref{app:Drinfeld}; here we consider the Moyal--Weyl twist \eqref{eq:MWtwist1}. Following \cite{Giotopoulos:2021ieg,UsQED}, we define the braided $L_\infty$-algebra for $\phi^4$-theory. The underlying graded vector space $V=V_1\oplus V_2$ and the kinetic bracket (\ref{eq:l1}) are unchanged, $\ell^\star_1 = \ell_1$, while the interaction bracket (\ref{eq:intbrackets}) changes to
\begin{align}
\ell_3^\star(\phi_1,\phi_2,\phi_3) = \lambda \, \phi_1\star\phi_2\star\phi_3 \ . \label{eq:intbracketsStar}
\end{align}
The Moyal--Weyl twist is compatible with the cyclic inner product \eqref{eq:scalarpairing}, $\langle-,-\rangle_\star=\langle-,-\rangle$, so the free braided scalar field theory is unchanged from its commutative version. 

The action functional for the interacting braided scalar field theory changes to
\begin{align} \label{SBraidedPhi4}
\begin{split}
S_{\star} &= \frac{1}{2!}\,\langle\phi,\ell_1(\phi)\rangle_\star  -\frac{1}{4!}\,\langle\phi,\ell^\star_3(\phi,\phi,\phi)\rangle_\star \\[4pt]
&= \int\d^4 x\ \left(\frac{1}{2}\,\phi\,\big(-\Box - m^2\big)\,\phi - \frac{\lambda}{4!}\,\phi\star\phi\star\phi\star\phi \right) \ .
\end{split}
\end{align}
At the classical level, it follows that braided scalar field theory is identical to the standard noncommutative scalar field theory.

\subsection{BV quantization of braided scalar field theory}
\label{sub:BVscalar}

Let us now explain how to compute correlation functions of the interacting braided scalar field theory using the braided BV formalism developed in \cite{SzaboAlex}. More details can be found in~\cite{UsQED}.

We start from the cohomology $H^\bullet(V)$ of the abelian $L_\infty$-algebra $(V,\ell_1)$, which describes the classical vacua of the free (braided) scalar field theory on $\FR^{1,3}$. This is also an abelian $L_\infty$-algebra concentrated in degrees~$1$ and~$2$, given by the solution space $H^1(V)=\ker(\ell_1)$ of the massive Klein--Gordon equation $(\square + m^2)\,\phi=0$ and the space $H^2(V)={\rm coker}(\ell_1)$, with the trivial differential~$0$.

Following the discussion in Section~\ref{sec:BVreview} and in Appendix~\ref{app:Drinfeld}, we need to define a translation-equivariant projection $\sfp: V\to H^\bullet(V)$ of degree $0$ and a translation-invariant contracting homotopy $\sfh: V_2\to V_1$. For this, let $\sgreen:C^\infty(\FR^{1,3})\to C^\infty(\FR^{1,3})$ denote the scalar Feynman propagator
\begin{align*}
\sgreen = -\frac1{\square+m^2} \qquad \mbox{with} \quad \tilde \sgreen(k) = \frac1{k^2-m^2} \ ,
\end{align*}
where $\tilde \sgreen(k)$ are the eigenvalues of the Green operator $\sgreen$ when acting on plane wave eigenfunctions of the form $\e^{\,\ii\, k\cdot x}$. It satisfies
\begin{align*}
\ell_1\circ \sgreen = -\big(\square+m^2\big)\circ \sgreen = \id_{C^\infty(\FR^{1,3})} \ .
\end{align*}

Since we are interested in calculating (braided) correlation functions, we take the trivial projection map $\sfp=0$,
or more accurately we restrict the cochain complex of $H^\bullet(V)$ to its trivial subspaces.
With these choices, the contracting homotopy $\sfh:V_2\to V_1$ is given by the propagator $\sfh = \sgreen$. Explicitly
\begin{align}\label{eq:htwo}
\sfh(\phi^+)(x) = -\frac1{\square+m^2}\, \phi^+(x) = \int \d^4 y \ \int_k \, \frac{\e^{-\ii\,k\cdot(x-y)}}{k^2-m^2} \ \phi^+(y) \ ,
\end{align}
for $\phi^+\in V_2$, where we use the shorthand $\int_k=\int\frac{\d^4k}{(2\pi)^4}$. In momentum space representation the contracting homotopy acts as
\begin{equation}
\sfh(\phi^+)(k) = \frac{\phi^+(k)}{k^2-m^2} .\nn
\end{equation}

Now we apply braided homological perturbation theory. As in the undeformed case, we extend the maps $\sfp$ and $\sfh$ to the braided space of functionals $\Sym_\RR (V[2])$ on $V$. The data above induce a trivial projection map $\sP:\Sym_\RR( V[2])\to \Sym_\RR( H^\bullet(V[2]))$ given by
\begin{align*}
\sP(1)=1 \qquad \mbox{and} \qquad \sP(\varphi_1\odot_\star\cdots\odot_\star\varphi_n) = 0 \ .
\end{align*}
The extended contracting homotopy $\sH:\Sym_\RR( V[2])\to\Sym_\RR( V[2])$ is defined as
\begin{align}\label{eq:sfH}
\begin{split}
\sH(1)&=0 \ , \\[4pt]
\sH(\varphi_1\odot_\star\cdots\odot_\star\varphi_n) &= \frac1n \, \sum_{a=1}^n \, \pm \ \varphi_1\odot_\star\cdots\odot_\star\varphi_{a-1}\odot_\star\sfh(\varphi_a) \odot_\star\varphi_{a+1} \odot_\star\cdots\odot_\star \varphi_n \ ,
\end{split}
\end{align}
for all $\varphi_1,\dots,\varphi_n\in V[2]$. We used the translation-invariance of $\sfh$ in (\ref{eq:sfH}) which trivializes the action of $\RR$-matrices. Note that on generators $\varphi_a \in V[2]$, the twisted symmetric product $\odot_\star$ is braided graded commutative:
\begin{align*}
\varphi_a\odot_\star\varphi_b = (-1)^{|\varphi_a|\,|\varphi_b|} \ \sfR_\alpha(\varphi_b)\odot_\star\sfR^\alpha(\varphi_a) \ .
\end{align*}

As in the undeformed case, we perturb the free differential $\ell_1$, extended as a graded derivation to all of $\Sym_\RR (V[2])$, to the `quantum' differential
\begin{align*}
Q_\BV = \ell_1 + \delta
\end{align*}
on $\Sym_\RR(V[2])$, with a formal translation-invariant perturbation $\delta$. The braided extension of the homological perturbation lemma~\cite{SzaboAlex} then constructs the perturbed  projection map $\tilde{\sP}=\sP+\sP_{\delta}$, where 
\begin{align*}
\sP_{\delta} = \sP\,\big(\id_{\Sym_\RR(V[2])} - {\delta}\,\sH\big)^{-1} \, \delta \, \sH \ .
\end{align*}
We are interested in two perturbations $\delta$ of $\ell_1$.

\paragraph{Free theory.}
Correlation functions for the free braided scalar field theory are calculated from the perturbation
\begin{align*}
{\delta} = \ii\,\hbar\,\BVL \ ,
\end{align*}
where $\BVL:\Sym_\RR (V[2])\to \Sym_\RR( V[2])[1]$ is the braided BV Laplacian defined by 
\begin{align}\label{eq:BVL}
\BVL(1)=0 & \qquad , \qquad \BVL(\varphi_1)=0 \qquad , \qquad  \BVL(\varphi_1\odot_\star\varphi_2) = \langle\varphi_1,\varphi_2\rangle_\star \ , \nn\\[4pt]
\BVL(\varphi_1\odot_\star\cdots\odot_\star\varphi_n) & = \sum_{a<b} \, \pm \, \langle \varphi_a,\sfR_{\alpha_{a+1}}\cdots\sfR_{\alpha_{b-1}}(\varphi_b)\rangle_\star \  \varphi_1\odot_\star\cdots\odot_\star\varphi_{a-1}\\ 
& \hspace{1.2cm} \odot_\star\sfR^{\alpha_{a+1}}(\varphi_{a+1})\odot_\star\cdots\odot_\star\sfR^{\alpha_{b-1}}(\varphi_{b-1})\odot_\star \varphi_{b+1}\odot_\star\cdots\odot_\star\varphi_n \ ,\nn
\end{align}
for all $\varphi_1,\dots,\varphi_n\in V[2]$. 

Since the braided BV Laplacian contracts fields pairwise and lowers the symmetric algebra degree from $n$ to $n-2$, it is clear that in this case the correlation functions \eqref{eq:BQFTnpoint} vanish unless $n=2k$ is even, in which case the free braided $2k$-point functions are then defined by
\begin{align}\label{eq:freeBQFT2k}
\begin{split}
G_{2k}^\star(x_1,\dots,x_{2k})^{(0)} &= \langle0|{\rm T}[\phi(x_1)\star\cdots\star\phi(x_{2k})]|0\rangle_\star^{(0)} \\[4pt]
:\!&= (\ii\,\hbar\,\BVL\,\sH)^k(\delta_{x_1}\odot_\star\cdots\odot_\star\delta_{x_{2k}}) \ .
\end{split}
\end{align}
Using \eqref{eq:scalarpairing}, \eqref{eq:htwo},  \eqref{eq:sfH} and \eqref{eq:BVL} we obtain the braided Wick expansion
\begin{align*}
\begin{split}
\ii\,\hbar\,\BVL\,\sH(\delta_{x_1}\odot_\star\cdots\odot_\star\delta_{x_{2k}}) &= \frac1{2k} \, \sum_{a<b} \, \bcontraction{}{\phi_a}{}{\sfR_{\alpha_{a+1}}\cdots\sfR_{\alpha_{b-1}}(\phi_b)} \phi_a\,\sfR_{\alpha_{a+1}}\cdots\sfR_{\alpha_{b-1}}(\phi_b) \  \delta_{x_1}\odot_\star\cdots\odot_\star\delta_{x_{a-1}} \\ & \quad  \odot_\star\sfR^{\alpha_{a+1}}(\delta_{x_{a+1}})\odot_\star\cdots\odot_\star\sfR^{\alpha_{b-1}}(\delta_{x_{b-1}})\odot_\star \delta_{x_{b+1}}\odot_\star\cdots\odot_\star\delta_{x_{2k}} \ .
\end{split}
\end{align*}

We use the notation 
\begin{align*}
\bcontraction{}{\phi_a}{}{\phi_b} \phi_a\,\phi_b:=\langle0| {\rm T}[\phi(x_a) \, \phi(x_b) ]|0\rangle^{(0)}=-\ii\,\hbar\,\sgreen(x_a-x_b)
\end{align*} 
for the free propagator and the fact that the only non-zero pairings are $\langle\delta_{x_a},\sgreen(\delta_{x_b})\rangle_\star = \sgreen(x_a-x_b)$. Note that the Koszul sign factors are trivial for antifields $\varphi_a=\delta_{x_a}\in V[2]_2$. Let us illustrate how these definitions work on the examples of two-point and four-point functions.

For the two-point function one finds immediately the free propagator
\begin{align*}
G^\star_2(x_1,x_2)^{(0)} &= \ii\,\hbar\,\BVL\,\sH(\delta_{x_1}\odot_\star\delta_{x_2})\\[4pt]
&= \ii\,\hbar\,\BVL\, \frac{1}{2}\,\big( \sgreen(\delta_{x_1})\odot_\star\delta_{x_2} + \delta_{x_1}\odot_\star \sgreen(\delta_{x_2}) \big)\\[4pt]
&= \ii\,\hbar\, \frac{1}{2}\,\big( \langle\sgreen(\delta_{x_1}), \delta_{x_2}\rangle_\star + \langle\delta_{x_1}, \sgreen(\delta_{x_2})\rangle_\star \big)\\[4pt]
&=  -\ii\,\hbar\, \int_k \, \frac{\e^{-\ii\,k\cdot(x_1-x_2)}}{k^2-m^2} = -\ii\,\hbar\, \sgreen(x_1-x_2) \ ,
\end{align*}
where we used (\ref{eq:htwo}) to calculate $\sgreen(\delta_{x_{a}})$. In momentum space representation, the two-point function is defined as
\begin{equation}
\tilde G^\star_2(k_1,k_2)^{(0)} = \ii\,\hbar\,\BVL\,\sH( \tte^{k_1}\odot_\star \tte^{k_2}) = \ii\,\hbar\, \frac{(2\pi)^4\,\delta(k_1+k_2)}{k^2_1-m^2} = \bcontraction{}{\phi_1}{}{\phi_2} \phi_1\,\phi_2 \ , \nn
\end{equation}
where $\tte^k(x) = \e^{\, \ii\,k\cdot x}$ and the contraction is done in momentum space.

For the four-point function in position space representation we start from
\begin{align*}
\begin{split}
\ii\,\hbar\,\BVL\,\sH(\delta_{x_1}\odot_\star\cdots\odot_\star\delta_{x_{4}}) &= \frac12\, \Big(\bcontraction{}{\phi_1}{}{\phi_2} \phi_1\,\phi_2\ \ \delta_{x_3}\odot_\star\delta_{x_4} + \bcontraction{}{\phi_1}{}{\sfR_\alpha(\phi_3)} \phi_1\,\sfR_\alpha(\phi_3) \ \sfR^\alpha(\delta_{x_2})\odot_\star\delta_{x_4} \\ & \hspace{1cm} + \bcontraction{}{\phi_1}{}{\sfR_\alpha\,\sfR_\beta(\phi_4)} \phi_1\,\sfR_\alpha\,\sfR_\beta(\phi_4) \ \sfR^\alpha(\delta_{x_2})\odot_\star\sfR^\beta(\delta_{x_3}) + \bcontraction{}{\phi_2}{}{\phi_3} \phi_2\,\phi_3 \ \delta_{x_1}\odot_\star\delta_{x_4} \\
& \hspace{1cm} + \bcontraction{}{\phi_2}{}{\sfR_\alpha(\phi_4)} \phi_2\,\sfR_\alpha(\phi_4) \ \delta_{x_1}\odot_\star\sfR^\alpha(\delta_{x_3}) + \bcontraction{}{\phi_3}{}{\phi_4} \phi_3\,\phi_4 \ \delta_{x_1}\odot_\star\delta_{x_2} \Big) \ .
\end{split}
\end{align*}
Using
\begin{align*}
\ii\,\hbar\,\BVL\,\sH(\delta_{x_a}\odot_\star\delta_{x_b}) = \bcontraction{}{\phi_a}{}{\phi_b} \phi_a\,\phi_b
\end{align*}
we then find that the four-point function is given by
\begin{align*}
\begin{split}
G_4^\star(x_1,x_2,x_3,x_4)^{(0)} &= (\ii\,\hbar\,\BVL\,\sH)^2 \, (\delta_{x_1}\odot_\star\delta_{x_2}\odot_\star\delta_{x_3}\odot_\star\delta_{x_4})  \\[4pt]
&= \frac12\,\Big(2\,\bcontraction{}{\phi_1}{}{\phi_2} \phi_1\,\phi_2\,\bcontraction{}{\phi_3}{}{\phi_4} \phi_3\,\phi_4 + \bcontraction{}{\phi_1}{}{\sfR_\alpha(\phi_3)} \phi_1\,\sfR_\alpha(\phi_3) \, \bcontraction{}{\sfR^\alpha(\phi_2)}{}{\phi_4} \sfR^\alpha(\phi_2)\,\phi_4 + \bcontraction{}{\phi_1}{}{\sfR^\alpha(\phi_3)} \phi_1\,\sfR^\alpha(\phi_3)\,\bcontraction{}{\phi_2}{}{\sfR_\alpha(\phi_4)} \phi_2\,\sfR_\alpha(\phi_4) \\
& \hspace{1cm} + \bcontraction{}{\phi_1}{}{\sfR_\alpha\,\sfR_\beta(\phi_4)} \phi_1\,\sfR_\alpha\,\sfR_\beta(\phi_4) \, \bcontraction{}{\sfR^\alpha(\phi_2)}{}{\sfR^\beta(\phi_3)} \sfR^\alpha(\phi_2)\,\sfR^\beta(\phi_3) + \bcontraction{}{\phi_1}{}{\phi_4} \phi_1\,\phi_4 \, \bcontraction{}{\phi_2}{}{\phi_3} \phi_2\,\phi_3 \Big) \ .
\end{split}
\end{align*}

We now employ some $\RR$-matrix manipulations using
\begin{align*}
\partial_i^2\big(\bcontraction{}{\phi_2}{}{\phi_4} \phi_2\,\phi_4\big) = -\partial_i^4\big(\bcontraction{}{\phi_2}{}{\phi_4} \phi_2\,\phi_4\big)
\end{align*}
to conclude that the second and third terms are equal. From the identity~\cite{UsQED}
\begin{align}\label{eq:4ptid}
\begin{split}
\bcontraction{}{\phi_1}{}{\sfR_\alpha \, \sfR_\beta(\phi_4)} \phi_1\, \sfR_\alpha \, \sfR_\beta(\phi_4)\, \bcontraction{}{\sfR^\alpha(\phi_2)}{}{\sfR^\beta(\phi_3)} \sfR^\alpha(\phi_2)\, \sfR^\beta(\phi_3) 
= \bcontraction{}{\phi_1}{}{\phi_4} \phi_1\, \phi_4\,\bcontraction{}{\phi_2}{}{\phi_3} \phi_2\, \phi_3 \ ,
\end{split}
\end{align}
we see that the last two terms are also equal. Altogether we arrive at
\begin{align}
G_4^\star(x_1,x_2,x_3,x_4)^{(0)} = \bcontraction{}{\phi_1}{}{\phi_2} \phi_1\,\phi_2\,\bcontraction{}{\phi_3}{}{\phi_4} \phi_3\,\phi_4 + \bcontraction{}{\phi_1}{}{\sfR_\alpha(\phi_3)} \phi_1\,\sfR_\alpha(\phi_3) \, \bcontraction{}{\sfR^\alpha(\phi_2)}{}{\phi_4} \sfR^\alpha(\phi_2)\,\phi_4 + \bcontraction{}{\phi_1}{}{\phi_4} \phi_1\,\phi_4 \, \bcontraction{}{\phi_2}{}{\phi_3} \phi_2\,\phi_3 \ . \label{4PointFull}
\end{align}

Repeating this calculation in momentum space results in
\begin{align}
\tilde{G}_4^\star(k_1,k_2,k_3,k_4)^{(0)} &= (\ii\,\hbar\,\BVL\,\sH)^2 \, (\tte^{k_1}\odot_\star \tte^{k_2}\odot_\star \tte^{k_3}\odot_\star \tte^{k_4})  \nn\\[4pt]
&= (\ii\,\hbar)^2 \, \Big( \frac{(2\pi)^4\,\delta(k_1+k_2)}{k^2_1-m^2} \, \frac{(2\pi)^4\,\delta(k_3+k_4)}{k^2_3-m^2} \nn\\
& \hspace{2cm}+ \frac{(2\pi)^4\,\delta(k_1+k_4)}{k^2_1-m^2} \,  \frac{(2\pi)^4\,\delta(k_2+k_3)}{k^2_2-m^2} \label{4PointFullMomentum} \\
& \hspace{3cm}+ \e^{\,\ii\,k_3\cdot\theta\,k_2}\,\frac{(2\pi)^4\,\delta(k_1+k_3)}{k^2_1-m^2} \, \frac{(2\pi)^4\,\delta(k_2+k_4)}{k^2_2-m^2}\Big) \nn\\[4pt]
&= \bcontraction{}{\phi_1}{}{\phi_2} \phi_1\,\phi_2 \, \bcontraction{}{\phi_3}{}{\phi_4} \phi_3\,\phi_4
+ \bcontraction{}{\phi_1}{}{\phi_4} \phi_1\,\phi_4 \, \bcontraction{}{\phi_2}{}{\phi_3} \phi_2\,\phi_3
+ \e^{\,\ii\,k_3\cdot\theta\,k_2} \, \bcontraction{}{\phi_1}{}{\phi_3} \phi_1\,\phi_3 \, \bcontraction{}{\phi_2}{}{\phi_4} \phi_2\,\phi_4 \  , \nn
\end{align}
where $k\cdot\theta\, p:=k_\mu\,\theta^{\mu\nu}\,p_\nu=-p\cdot\theta\, k$. The last term in (\ref{4PointFullMomentum}) is exactly the term $\bcontraction{}{\phi_1}{}{\sfR_\alpha(\phi_3)} \phi_1\,\sfR_\alpha(\phi_3) \, \bcontraction{}{\sfR^\alpha(\phi_2)}{}{\phi_4} \sfR^\alpha(\phi_2)\,\phi_4$ in (\ref{4PointFull}).

\paragraph{Interacting theory.} 
An interacting braided scalar quantum field theory is captured by the perturbation
\begin{align}
\delta = \ii\,\hbar\,\BVL + \{\CS _{\rm int},-\}_\star \ . \label{Pert2}
\end{align}
 The antibracket in (\ref{Pert2}) is the braided graded Poisson bracket $$\{-,-\}_\star:\Sym_\RR( V[2])\otimes \Sym_\RR (V[2])\longrightarrow \Sym_\RR (V[2])[1]$$ defined by setting 
\begin{align*}
\{\varphi_a,\varphi_b\}_\star = \langle\varphi_a , \varphi_b\rangle_\star =\pm\,\{\sfR_\alpha(\varphi_b),\sfR^\alpha(\varphi_a)\}_\star
\end{align*}
for $\varphi_a\in V[2]$, and extending this to all of $\Sym_\RR (V[2])$ as a braided graded Lie bracket which is a braided graded derivation on $\Sym_\RR (V[2])$ in each of its slots; for example
\begin{align}\label{eq:braidedder}
\{\varphi_1,\varphi_2\odot_\star\varphi_3\}_\star = \langle\varphi_1,\varphi_2\rangle_\star\odot_\star\varphi_3 \pm \sfR_\alpha(\varphi_2)\odot_\star\langle\sfR^\alpha(\varphi_1),\varphi_3\rangle_\star \ .
\end{align}

As in the undeformed case, the antibracket is compatible with the differential $\ell_1$, as a consequence of cyclicity of the inner product $\langle-,-\rangle_\star$. It is also related to the braided BV Laplacian through
\begin{align}\label{eq:BVLantibracketBraided}
\BVL(a_1\odot_\star a_2) = \BVL(a_1)\odot_\star a_2 + (-1)^{\vert a_1\vert}\, 
a_1\odot_\star\BVL(a_2) +  \{a_1,a_2\}_\star \ ,
\end{align}
for all $a_1,a_2\in \Sym_\RR( L[2])$.
 
For $\lambda\,\phi^4$-theory on $\FR^{1,3}$, the interaction action $\CS _{\rm int}\in \Sym_\RR(V[2])$ in (\ref{Pert2}) is defined by
\begin{align}\label{eq:Sintxi}
\CS _{\rm int} := -\frac1{4!} \, \langle \xi \,,\,\ell^{\star \, {\rm ext}}_3(\xi,\xi,\xi)\rangle^{\rm ext}_\star \ .
\end{align}
The contracted coordinate functions $\xi\in\Sym_\RR (L[2])\otimes V$ are given by
\begin{align*}
\xi = \int_k \, \big(\tte_k\otimes \tte^k + \tte^k\otimes \tte_k\big) \ ,
\end{align*}
where $\tte_k(x)=\e^{-\ii\, k\cdot x}$ is the basis of plane waves for $V_1$ with dual basis 
$
\tte^k(x)=\tte_k^*(x)=\e^{\,\ii\,k\cdot x}
$
for~$V_2$. 

These bases are dual with respect to the inner product \eqref{eq:scalarpairing}, in the sense that
\begin{align*}
\int_p \ \langle \tte_k,\tte^p\rangle_\star \ \tte_p = \tte_k \qquad \mbox{and} \qquad \int_k \ \tte^k \ \langle \tte_k,\tte^p\rangle_\star = \tte^p \ ,
\end{align*}
where throughout we use
\begin{align*}
\int \dd^4x \ \e^{\pm\,\ii\,k\cdot x} = (2\pi)^4 \, \delta(k) \ .
\end{align*}
The star-products among basis fields are
\begin{align}\label{eq:ekstarep}
\tte_k\star \tte_p = \e^{\, -\frac\ii2 \, k\cdot\theta\, p} \ \tte_{k+p} \ ,
\end{align}
while the action of the inverse $\RR$-matrix on them is given by
\begin{align}\label{eq:Rekotimesep}
\RR^{-1}(\tte_k\otimes \tte_p) = \sfR_\alpha(\tte_k)\otimes\sfR^\alpha(\tte_p) = \e^{\,\ii\,k\cdot\theta\,p} \ \tte_k\otimes \tte_p \ .
\end{align}

As discussed in~\cite{SzaboAlex,Giotopoulos:2021ieg}, the interaction action (\ref{eq:Sintxi}) satisfies the classical master equation
\begin{align*}
\ell_1(\CS _{\rm int}) + \frac12\,\{\CS _{\rm int},\CS _{\rm int}\}_\star=0 \ , 
\end{align*}
and it is annihilated by the braided BV Laplacian, $\BVL(\CS _{\rm int})=0$. As a consequence, the operator 
\begin{align*}
Q_{\BV}=\ell_1 + \ii\,\hbar\,\BVL + \{\CS _{\rm int},-\}_{\star}
\end{align*}
is a differential, $(Q_{\BV})^2=0$, which describes the correlation functions in terms of a braided quantum $L_\infty$-algebra.

To explicitly calculate (\ref{eq:Sintxi}), we extend the braided $L_\infty$-structure on $V$ to $\Sym_\RR( V[2])\otimes V$ via the non-zero brackets~\cite{Giotopoulos:2021ieg}
\begin{align}
\ell^{\rm{ext}}_1(a_1\otimes\phi_1) &= \pm\,a_1\otimes\ell_1(\phi_1) \ ,\nn \\[4pt]
\ell_3^{\star\, {\rm{ext}}}(a_1\otimes\phi_1,a_2\otimes\phi_2,a_3\otimes\phi_3) &= \pm\,\big(a_1\odot_\star\sfR_\alpha(a_2)\odot_\star\sfR_\beta\,\sfR_\sigma(a_3)\big)\nn\\
& \hspace{2cm} \otimes\ell_3^\star\big(\sfR^\beta\,\sfR^\alpha(\phi_1),\sfR^\sigma(\phi_2),\phi_3\big) \ ,\nn
\end{align}
for $a_1,a_2,a_3\in\Sym_\RR (V[2])$ and $\phi_1,\phi_2,\phi_3\in V_1$; again we write $\pm$ for the Koszul sign factors determined by the gradings of the elements involved in all operations. Similarly, the cyclic structure is extended via the non-zero $\Sym_\RR (V[2])$-valued pairing
\begin{align}\nn
\langle a_1\otimes\phi,a_2\otimes\phi^+\rangle^{\rm ext}_\star = \pm\, \big(a_1\odot_\star\sfR_\alpha(a_2)\big) \, \langle\sfR^\alpha(\phi),\phi^+\rangle_\star \ ,
\end{align}
for $a_1,a_2\in\Sym_\RR (V[2])$, $\phi\in V_1$ and $\phi^+\in V_2$.

Explicit calculation of \eqref{eq:Sintxi} using \eqref{eq:scalarpairing}, \eqref{eq:ekstarep} and \eqref{eq:Rekotimesep} results in
\begin{align*}
\begin{split}
\CS _{\rm int}&= -\frac1{4!} \, \int_{k_1,k_2,k_3,k_4} \, \langle \tte^{k_1}\otimes \tte_{k_1}\,,\,\ell^{\star\,\rm{ext}}_3(\tte^{k_2}\otimes \tte_{k_2},\tte^{k_3}\otimes \tte_{k_3},\tte^{k_4}\otimes \tte_{k_4})\rangle^{\rm{ext}}_\star \\[4pt]
&= -\frac1{4!} \, \int_{k_1,k_2,k_3,k_4} \, \langle \tte^{k_1}\otimes \tte_{k_1}\,,\, \big(\tte^{k_2}\odot_\star\sfR_\alpha(\tte^{k_3})\odot_\star \sfR_\beta\,\sfR_\sigma(\tte^{k_4})\big) \\
& \hspace{6cm}\otimes\ell_3^\star\big(\sfR^\beta\,\sfR^\alpha(\tte_{k_2}),\sfR^\sigma(\tte_{k_3}),\tte_{k_4}\big)\rangle^{\rm{ext}}_\star \\[4pt]
&= -\frac1{4!} \, \int_{k_1,k_2,k_3,k_4} \, \e^{\,\ii\,k_2\cdot\theta\, k_3+\ii\,k_2\cdot\theta\, k_4 + \ii\,k_3\cdot\theta\, k_4} \\
& \hspace{4cm} \times \langle \tte^{k_1}\otimes \tte_{k_1} \,,\, (\tte^{k_2}\odot_\star \tte^{k_3}\odot_\star \tte^{k_4}) \otimes \ell_3^\star(\tte_{k_2},\tte_{k_3},\tte_{k_4})\rangle^{\rm{ext}}_\star \\[4pt]
&= -\frac1{4!} \, \int_{k_1,k_2,k_3,k_4} \, \e^{\,\ii\,k_2\cdot\theta\, k_3+\ii\,k_2\cdot\theta\, k_4 + \ii\,k_3\cdot\theta\, k_4}\\
& \hspace{4cm} \times \tte^{k_1}\odot_\star\sfR_\alpha(\tte^{k_2}\odot_\star \tte^{k_3}\odot_\star \tte^{k_4}) \, \langle\sfR^\alpha(\tte_{k_1}),\ell_3^\star(\tte_{k_2},\tte_{k_3},\tte_{k_4})\rangle_\star \\[4pt]
&= -\frac\lambda{4!} \, \int_{k_1,k_2,k_3,k_4} \, \e^{\,\ii\,\sum\limits_{a<b} \, k_a\cdot\theta\, k_b} \ \tte^{k_1}\odot_\star \tte^{k_2}\odot_\star \tte^{k_3}\odot_\star \tte^{k_4} \ \langle \tte_{k_1},\tte_{k_2}\star \tte_{k_3}\star \tte_{k_4}\rangle_\star \\[4pt]
&=-\frac\lambda{4!} \, \int_{k_1,k_2,k_3,k_4} \, \e^{\,\ii\,\sum\limits_{a<b} \, k_a\cdot\theta\, k_b} \ \e^{-\,\frac\ii2\,\sum\limits_{a<b} \, k_a\cdot\theta\, k_b}\\
& \hspace{4cm} \times (2\pi)^4 \, \delta(k_1+k_2+k_3+k_4) \ \tte^{k_1}\odot_\star \tte^{k_2}\odot_\star \tte^{k_3}\odot_\star \tte^{k_4}\\[4pt]
&=:  \int_{k_1,k_2,k_3,k_4} \, V_4(k_1,k_2,k_3,k_4) \ \tte^{k_1}\odot_\star \tte^{k_2}\odot_\star \tte^{k_3}\odot_\star \tte^{k_4}  \ .
\end{split}
\end{align*}

The interaction vertex 
\begin{align}\label{eq:Vint}
V_4(k_1,k_2,k_3,k_4) = -\frac\lambda{4!} \ \e^{\,\frac\ii2\,\sum\limits_{a<b} \, k_a\cdot\theta\, k_b} \ (2\pi)^4 \, \delta(k_1+k_2+k_3+k_4)
\end{align}
coincides with the vertex of the standard noncommutative $\lambda\,\phi_4^{4}$-theory~\cite{Minwalla:1999px,U1Reviewa}. It has the braided symmetry
\begin{align}\label{eq:Vbraidedsym}
V_4(\ \  k_{a+1},k_a\ \ ) = \e^{-\ii\,k_a\cdot\theta\, k_{a+1}} \ V_4(k_1,k_2,k_3,k_4)
\end{align}
under interchange of any pair of neighbouring momenta, and also the cyclic symmetry
\begin{align}\label{eq:Vcyclicsym}
V_4(k_1,k_2,k_3,k_4) = V_4(k_4,k_1,k_2,k_3)
\end{align}
which follows from momentum conservation.

The interacting correlation functions of the braided quantum field theory are now given by
\begin{align}\label{eq:intcorrelationfn}
\begin{split}
G_n^\star(x_1,\dots,x_n)^{\rm int}&=\langle 0|{\rm T}[\phi(x_1)\star\cdots\star\phi(x_n)]|0\rangle^{\rm int} \\[4pt]
:\!&= \sum_{m=1}^\infty \, \sP\,\big((\ii\,\hbar\,\BVL\,\sH + \{\CS _{\rm int},-\}_\star\,\sH)^m\,(\delta_{x_1}\odot_\star\cdots\odot_\star\delta_{x_n})\big) \ ,
\end{split}
\end{align}
or in  momentum space representation
\begin{equation}\label{eq:intcorrelationfnMomentum}
\tilde G_n^\star(p_1,\dots,p_n)^{\rm int} = \sum_{m=1}^\infty \, \sP\,\big((\ii\,\hbar\,\BVL\,\sH + \{\CS _{\rm int},-\}_\star\,\sH)^m\, (\tte^{p_1}\odot_\star\cdots\odot_\star\tte^{p_n})\big) \ .
\end{equation}
This is a formal power series expansion in $\hbar$ and the coupling constant $\lambda$.

\begin{figure}[h]%
\centering
\includegraphics[width=0.4\textwidth]{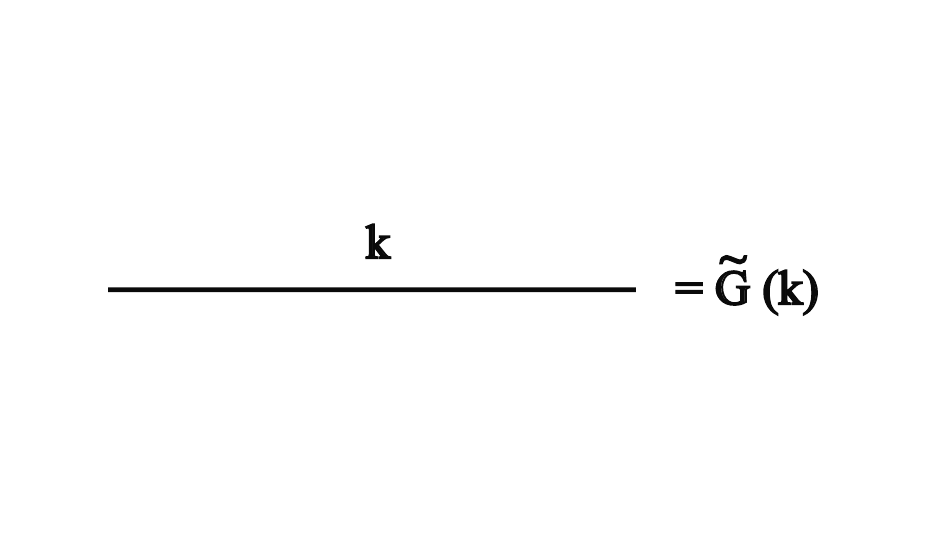} \qquad
\includegraphics[width=0.45\textwidth]{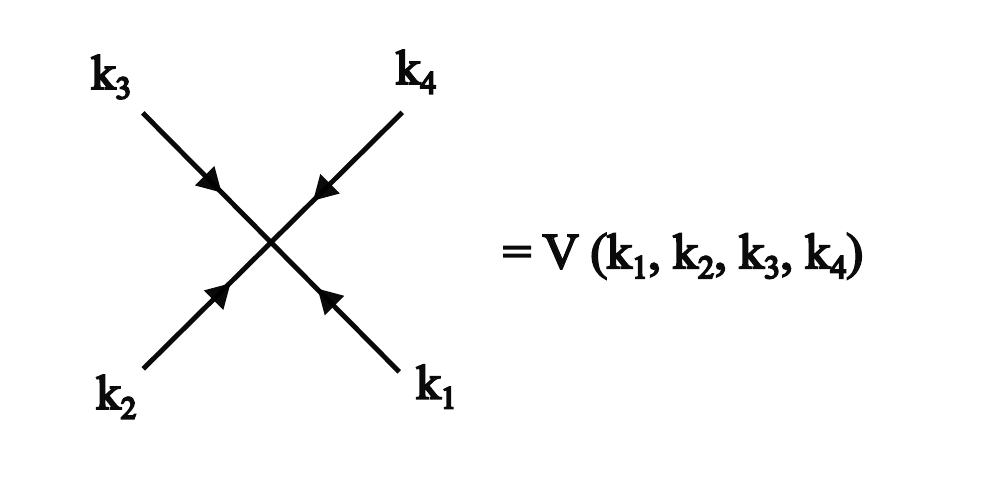}
\vspace{-5mm}
\caption{\small Diagrammatic representation of the propagator (left) and interaction vertex (right) for braided $\lambda\,\phi_4^4$-theory.
}\label{fig:srules}
\end{figure}

We represent terms in this perturbative expansion using standard Feynman diagrammatic techniques; the Feynman rules are depicted in Figure~\ref{fig:srules}.

\section{Examples: Two-point functions at one-loop}
\label{sec:examples}

In this final section we consider two illustrative examples: the two-point functions at one-loop order for braided $\phi^4$-theory in four dimensions and braided $\phi^3$-theory in six dimension. We will present all calculations in momentum space representation. 

\subsection{Braided $\phi^4$-theory}

We consider first the braided $\phi^4$-theory on $\FR^{1,3}$ from Section~\ref{sec:BQFTreview}, which was studied in~\cite{UsQED} in position space representation. Let us compute the  one-loop correction   to the free two-point function from Section~\ref{sub:BVscalar} in momentum space. Using \eqref{eq:intcorrelationfn}, this is given by 
\begin{equation}
\tilde G_2^\star(p_1,p_2)^{(1)} = (\ii\,\hbar\,\BVL\,\sH)^2\,\{\CS _{\rm int},\sH(\tte^{p_1}\odot_\star \tte^{p_2})\}_\star \ .  \label{eq:2pt1loopMomentum} 
\end{equation}

We first calculate $\{\CS _{\rm int},\sH(\tte^{p_1}\odot_\star \tte^{p_2})\}_\star$  using the non-zero pairings 
\begin{align}\label{eq:eGdeltapairMomentum}
\langle \tte^{k} , \sgreen(\tte^{p})\rangle_\star = \langle \sgreen(\tte^{p}) , \tte^{k}\rangle_\star = \langle \tte^{p} ,  \sgreen(\tte^{k})\rangle_\star = \tilde \sgreen(p) \, (2\pi)^4 \, \delta(k+p) \ ,
\end{align}
together with the braided derivation property \eqref{eq:braidedder} of the antibracket as well as the  symmetry properties \eqref{eq:Vbraidedsym} and \eqref{eq:Vcyclicsym} of the interaction vertex $V_4(k_1,k_2,k_3,k_4)$. We find
\begin{align*}
\{\CS _{\rm int},\sH(\tte^{p_1}\odot_\star \tte^{p_2})\}_\star & = \frac{4}{2} \, \int_{k_1,k_2,k_3,k_4} V_4(k_1,k_2,k_3,k_4) \, \bigl( \langle \tte^{k_4}, \sgreen(\tte^{p_1})\rangle_\star \,  \tte^{k_1}\odot_\star \tte^{k_2}\odot_\star \tte^{k_3}\odot_\star \tte^{p_2} \, \\
& \hspace{3.5cm} + \langle \tte^{k_4}, \sgreen(\tte^{p_2})\rangle_\star \,  \tte^{p_1}\odot_\star \tte^{k_1}\odot_\star \tte^{k_2}\odot_\star \tte^{k_3} \bigr)  \ .
\end{align*}
Inserting this  into (\ref{eq:2pt1loopMomentum}) gives
\begin{align}\label{eq:2pt1loop1Momentum} 
\begin{split}
\tilde G_2^\star(p_1,p_2)^{(1)}
&= \frac{4}{2} \, \int_{k_1,k_2,k_3,k_4} \, V_4(k_1,k_2,k_3,k_4) \\
& \hspace{2cm}\times \Bigl( \langle \tte^{k_4}, \sgreen(\tte^{p_1})\rangle_\star \,  (\ii\,\hbar\,\BVL\,\sH)^2\,\bigl(\tte^{k_1}\odot_\star \tte^{k_2}\odot_\star \tte^{k_3}\odot_\star \tte^{p_2} \bigr) \\
& \hspace{2.5cm} + \langle \tte^{k_4}, \sgreen(\tte^{p_2})\rangle_\star \,  (\ii\,\hbar\,\BVL\,\sH)^2\, \bigl(\tte^{p_1}\odot_\star \tte^{k_1}\odot_\star \tte^{k_2}\odot_\star \tte^{k_3} \bigr) \Bigr)\  .
\end{split}
\end{align}

The free four-point functions in \eqref{eq:2pt1loop1Momentum} are evaluated by using the braided Wick expansion \eqref{4PointFull} in momentum space. We obtain
\begin{align}\label{eq:G24pt1Momentum}
\begin{split}
& (\ii\,\hbar\,\BVL\,\sH)^2\, \big(\tte^{k_1}\odot_\star \tte^{k_2}\odot_\star \tte^{k_3}\odot_\star\tte^{p_2}\big) \\[4pt]
 & \hspace{1cm} = -\hbar^2 \, \big(\langle \tte^{k_1},\sgreen(\tte^{k_2})\rangle_\star \, \langle \tte^{k_3},\sgreen(\tte^{p_2})\rangle_\star \\
& \hspace{2.5cm}+ \e^{\,\ii\,k_3\cdot\theta\,k_2}\,\langle \tte^{k_1},\sgreen(\tte^{k_3})\rangle_\star \, \langle \tte^{k_2},\sgreen(\tte^{p_2})\rangle_\star  + \langle \tte^{k_1},\sgreen(\tte^{p_2})\rangle_\star \, \langle \tte^{k_2},\sgreen(\tte^{k_3})\rangle_\star\big) \\[4pt]
&\hspace{1cm}  = -\hbar^2 \, (2\pi)^8 \, \tilde \sgreen(p_2) \, \big(\delta(k_1+k_2)\,\delta(k_3+p_2) \, \tilde \sgreen(k_1)  \\
& \hspace{2.5cm} + \e^{\,\ii\,k_3\cdot\theta\,k_2}\, \delta(k_1+k_3)\,\delta(k_2+p_2) \, \tilde \sgreen(k_1) + \delta(k_1+p_2)\,\delta(k_2+k_3) \, \tilde \sgreen(k_2)\big) \ ,
\end{split}
\end{align}
and similarly
\begin{align}\label{eq:G24pt2Momentum}
\begin{split}
& \big(\ii\,\hbar\,\BVL\,\sH)^2\, \big(\tte^{p_1}\odot_\star \tte^{k_1}\odot_\star \tte^{k_2}\odot_\star \tte^{k_3}\big) \\[4pt] & \hspace{1cm} = -\hbar^2 \, (2\pi)^8 \, \tilde \sgreen(p_1) \, \big(\delta(k_1+p_1)\,\delta(k_3+k_2) \, \tilde \sgreen(k_2)  \\
& \hspace{2.5cm} + \e^{\,\ii\,k_2\cdot\theta\,k_1}\, \delta(p_1+k_2)\,\delta(k_1+k_3) \, \tilde \sgreen(k_1) + \delta(k_3+p_1)\,\delta(k_2+k_1) \, \tilde \sgreen(k_1)\big) \ .
\end{split}
\end{align}

We now substitute \eqref{eq:G24pt1Momentum} and \eqref{eq:G24pt2Momentum} into \eqref{eq:2pt1loop1Momentum}, and resolve the delta-functions. We find that all six contributions are the same. Altogether the one-loop contribution to the two-point function in momentum space is given by
\begin{align}\label{eq:2pt1loopfinalMomentum}
\tilde G_2^\star(p_1,p_2)^{(1)} = \frac{\hbar^2\,\lambda}{2} \, \frac{(2\pi)^4\, \delta(p_1+p_2)}{(p_1^2 - m^2)\,(p_2^2 - m^2)} \ \int_k\, \frac{1}{k^2 - m^2} \ .
\end{align}

\begin{figure}[h]%
\centering
\includegraphics[width=0.45\textwidth]{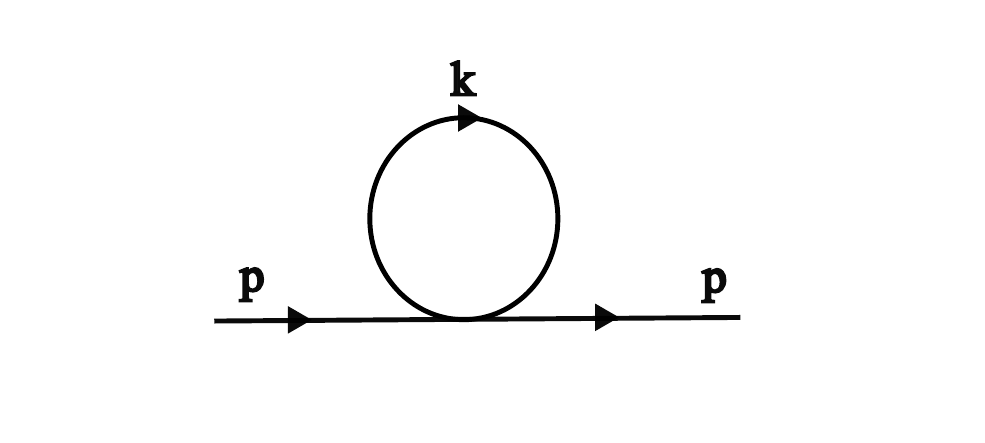} 
\vspace{-5mm}
\caption{\small In braided $\lambda\,\phi_4^4$-theory, the one-loop self-energy receives a contribution only from a planar tadpole diagram.}
\label{fig:stadpole}
\end{figure}

This result agrees with the position space calculation given in~\cite{UsQED}. It is independent of the deformation parameter and coincides with the classical two-point function (at $\theta=0$), including the correct sign and overall combinatorial factor. It shows that there is no UV/IR mixing in the two-point function at one-loop order, in contrast to the standard noncommutative quantum field theory~\cite{Minwalla:1999px,U1Reviewa}. It also suggests that there are no non-planar Feynman diagrams in perturbation theory, see Figure~\ref{fig:stadpole}. This appears to be a consequence of the braided symmetries of the interaction vertex due to the braided $L_\infty$-structure, through its interplay with the braided Wick theorem.  

\subsection{Braided $\phi^3$-theory}
\label{sub:braidedphi3}

We extend the example of braided $\lambda\,\phi^4$-theory in four dimensions to braided $\lambda\,\phi^3$-theory in six dimensions. This is based on the braided $L_\infty$-algebra with graded vector space $V=V_1\oplus V_2$, where $V_1=V_2=C^\infty(\FR^{1,5})$, differential $\ell_1^\star=\ell_1=-\square-m^2$, and the single non-vanishing higher bracket $\ell_2^\star$ defined as
\begin{align}
\ell_2^\star(\phi_1,\phi_2) = \lambda \, \phi_1\star\phi_2 \ ,\label{l2Phi3}
\end{align}
for $\lambda\in\FR$ and $\phi_1,\phi_2\in V_1$.

The calculation of the interaction action functional $\CS _{\rm int}\in\Sym_\RR(V[2])$ proceeds as in Section~\ref{sub:BVscalar}. We obtain
\begin{align*}
\begin{split}
\CS _{\rm int} & = -\frac1{3!} \, \int_{k_1,k_2,k_3} \, \langle \tte^{k_1}\otimes \tte_{k_1}\,,\,\ell_2^{\star\, {\rm ext}}(\tte^{k_2}\otimes \tte_{k_2},\tte^{k_3}\otimes \tte_{k_3})\rangle^{\rm ext}_\star \\[4pt]
& =: \int_{k_1,k_2,k_3} \, V_3(k_1,k_2,k_3) \ \tte^{k_1}\odot_\star \tte^{k_2}\odot_\star \tte^{k_3} \ ,
\end{split}
\end{align*}
where the interaction vertex is
\begin{align}\label{eq:Vint3}
V_3(k_1,k_2,k_3) = -\frac\lambda{3!} \ \e^{\,\frac\ii2\,\sum\limits_{a<b} \, k_a\cdot\theta\, k_b} \ (2\pi)^6 \, \delta(k_1+k_2+k_3) \ ,
\end{align}
with $a,b =1,2,3$. Analogously to (\ref{eq:Vbraidedsym}) and (\ref{eq:Vcyclicsym}), the vertex function (\ref{eq:Vint3}) has the braided symmetry
\begin{align}\label{eq:V3braidedsym}
V_3(\ \  k_{a+1},k_a\ \ ) = \e^{\, -\ii\,k_a\cdot\theta\, k_{a+1}} \ V_3(k_1,k_2,k_3)
\end{align}
under interchange of any pair of neighbouring momenta, and also the cyclic symmetry
\begin{align}\label{eq:V3cyclicsym}
V_3(k_1,k_2,k_3) = V_3(k_3,k_1,k_2) \ .
\end{align}

The two-point function is defined again as
\begin{equation}\label{eq:2PointPhi3}
\tilde G_2^\star(p_1,p_2)^{\rm int} = \sum_{m=1}^\infty \, \sP\,\big((\ii\,\hbar\,\BVL\,\sH + \{\CS _{\rm int},-\}_\star\,\sH)^m\, (\tte^{p_1}\odot_\star\tte^{p_2})\big) .
\end{equation}
The one-loop corrections to the free two-point function are given by the sum of two terms
\begin{align}\label{eq:2pt2loop}
\begin{split}
& \tilde G_2^\star(p_1,p_2)_1^{(1)} = \ii\,\hbar\,\BVL\,\sH \,\bigl\{ \CS _{\rm int},\sH\, \big((\ii\,\hbar\,\BVL\,\sH)\,\bigl\{\CS _{\rm int},\sH\,(\tte^{p_1}\odot_\star \tte^{p_2})\bigr\}_\star \big)\bigr\}_\star \ , \\[4pt]
& \tilde G_2^\star(p_1,p_2)_2^{(1)} = (\ii\,\hbar\,\BVL\,\sH)^2 \, \bigl\{ \CS _{\rm int},\sH\, \bigl\{\CS _{\rm int},\sH\,(\tte^{p_1}\odot_\star \tte^{p_2})\bigr\}_\star \bigr\}_\star \ .
\end{split}
\end{align}
We start with 
\begin{align}\label{S2}
\begin{split}
& \bigl\{\CS _{\rm int},\sH(\tte^{p_1}\odot_\star \tte^{p_2})\bigr\}_\star = \frac{3}{2} \, \int_{k_1,k_2,k_3} \, V_3(k_1,k_2,k_3) \, \bigl( \langle \tte^{k_3}, \sgreen(\tte^{p_1})\rangle_\star \,  \tte^{k_1}\odot_\star \tte^{k_2}\odot_\star \tte^{p_2} \\
& \hspace{8cm} + \langle \tte^{k_3},  \sgreen(\tte^{p_2})\rangle_\star \,  \tte^{p_1}\odot_\star \tte^{k_1}\odot_\star \tte^{k_2} \bigr) \ ,
\end{split}
\end{align}
where the non-zero pairings are of the form
\begin{align}\nn
\begin{split}
& \langle \tte^{k} , \sgreen (\tte^{p})\rangle_\star = \langle \sgreen(\tte^{p}) , \tte^{k}\rangle_\star = \langle \tte^{p} , \sgreen(\tte^{k})\rangle_\star = \tilde \sgreen(p) \ (2\pi)^6 \, \delta(k+p) \ .
\end{split}
\end{align}

To calculate $\tilde G_2^\star(p_1,p_2)_1^{(1)}$, we apply $\ii\,\hbar\,\BVL\,\sH$ to (\ref{S2}) and obtain
\begin{align}\nn
&(\ii\,\hbar\,\BVL\,\sH)\,\bigl\{\CS _{\rm int},\sH(\tte^{p_1}\odot_\star \tte^{p_2})\bigr\}_\star \\[4pt]
& \qquad = \frac{3}{2} \, \int_{k_1,k_2,k_3} \, V_3(k_1,k_2,k_3) \,
\bigl( \langle \tte^{k_3},\sgreen(\tte^{p_1})\rangle_\star \,  (\ii\,\hbar\,\BVL\,\sH) \, (\tte^{k_1}\odot_\star \tte^{k_2}\odot_\star \tte^{p_2}) \nn\\
& \hspace{6cm} + \langle \tte^{k_3}, \sgreen(\tte^{p_2})\rangle_\star \, (\ii\,\hbar\,\BVL\,\sH) \, (\tte^{p_1}\odot_\star \tte^{k_1}\odot_\star \tte^{k_2}) \bigr) \ .\nn
\end{align}
These two terms result in
\begin{align} \label{H3,1}
\begin{split}
(\ii\,\hbar\,\BVL\,\sH) \, (\tte^{k_1}\odot_\star \tte^{k_2}\odot_\star \tte^{p_2}) &= \frac{2}{3} \, \ii\,\hbar \, \bigl( \langle \tte^{k_1} , \sgreen(\tte^{k_2})\rangle_\star\, \tte^{p_2}\\ 
&\hspace{10mm} + \langle \tte^{k_2} , \sgreen(\tte^{p_2})\rangle_\star\, \tte^{k_1} \, + \, \e^{\,-\ii\,k_1\cdot\theta\,k_2} \, \langle \tte^{k_1} , \sgreen(\tte^{p_2})\rangle_\star\, \tte^{k_2}\bigr)
\end{split}
\end{align}
and
\begin{align} \label{H3,2}
\begin{split}
(\ii\,\hbar\,\BVL\,\sH) \, (\tte^{p_1}\odot_\star \tte^{k_1}\odot_\star \tte^{k_2}) &= \frac{2}{3} \, \ii\,\hbar \, \bigl( \langle \tte^{k_1} , \sgreen(\tte^{p_1})\rangle_\star\, \tte^{k_2}\\ 
&\hspace{10mm} +\langle \tte^{k_1} , \sgreen(\tte^{k_2})\rangle_\star\, \tte^{p_1} \, + \, \e^{\,\ii\,k_2\cdot\theta\,k_1} \, \langle \tte^{p_1} , \sgreen(\tte^{k_2})\rangle_\star\, \tte^{k_1}\bigr) \ . 
\end{split}
\end{align}

\begin{figure}[h]%
\centering
\includegraphics[scale=0.7]{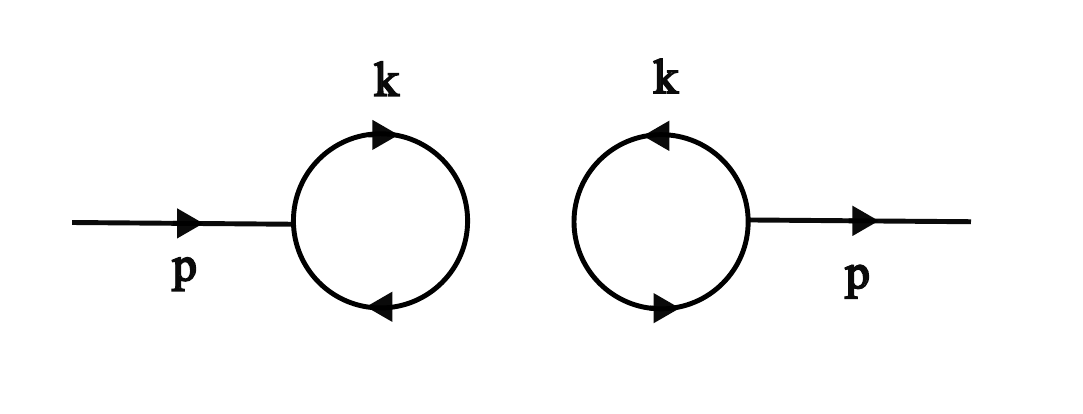} \qquad
\includegraphics[scale=0.7]{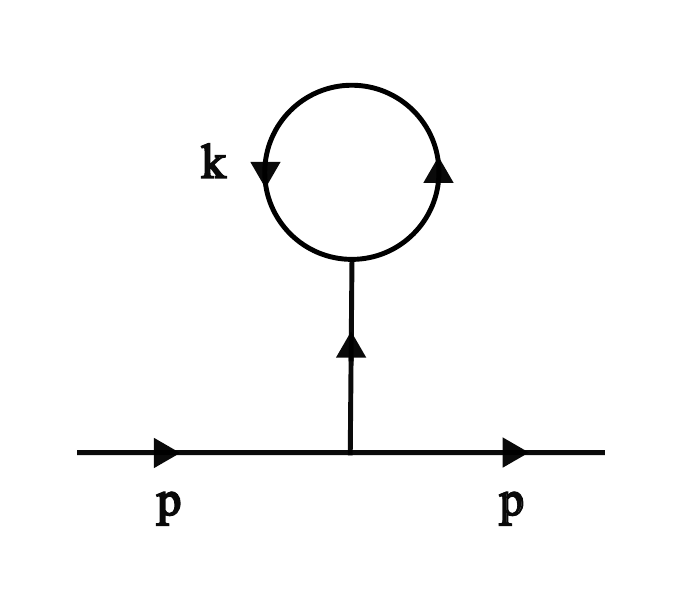}
\vspace{-5mm}
\caption{\small The tadpole contributions to the two-point function at one-loop in braided $\lambda\,\phi_6^3$-theory.}
\label{fig:Phi31}
\end{figure}

Adding another vertex insertion introduces three more internal momenta, two of which will be contracted via another application of $\ii\,\hbar\,\BVL\,\sH$, resulting in
\begin{align} \label{DeltaSH1}
\begin{split}
& (\ii\,\hbar\,\BVL\,\sH)\, \{\CS _{\rm int},\sH(\tte^{r})\}_\star = 3 \, \int_{q_1,q_2,q_3} \, V_3(q_1,q_2,q_3) \,  \langle \tte^{q_3}, \sgreen(\tte^{r})\rangle_\star \,  \langle \sgreen(\tte^{q_1}),\tte^{q_2}\rangle_\star \ ,
\end{split}
\end{align}
where we used the symmetry properties (\ref{eq:V3braidedsym}) and (\ref{eq:V3cyclicsym}). Applying (\ref{DeltaSH1}) to (\ref{H3,1}) and (\ref{H3,2}), and adding all terms, we get the first correction term
\begin{align} \label{PrvaKorekcija}
\begin{split}
\tilde G_2^\star(p_1,p_2)_1^{(1)} &= \frac{(\ii\,\hbar\,\lambda)^2}{6}\, \frac{(2\pi)^6\, \delta(p_1)}{p_1^2-m^2} \ \left[\int_{k}\, \frac{1}{k^2-m^2} \right]^2 \ \frac{(2\pi)^6\, \delta(p_2)}{p_2^2-m^2} \\[4pt]
& \quad\, +\frac{(\ii\,\hbar\,\lambda)^2}{3}\, \frac{(2\pi)^6\, \delta(p_1+p_2)}{(p_1^2 - m^2)\,(p_2^2 - m^2)} \ \int_k\, \frac{1}{(0-m^2)\,(k^2 - m^2)} \ .
\end{split}
\end{align}
We recognise the first term in $\tilde G_2^\star(p_1,p_2)_1^{(1)}$ as the contribution from the disconnected pair of tadpole diagrams and the second term as the contribution from the connected tadpole diagram, see Figure~\ref{fig:Phi31}.

To calculate $\tilde G_2^\star(p_1,p_2)_2^{(1)}$, we first have to add one more vertex to (\ref{S2}):
\begin{align}
& \bigl\{ \CS _{\rm int},\sH\, \{\CS _{\rm int},\sH\,(\tte^{p_1}\odot_\star \tte^{p_2})\}_\star \bigr\}_\star \nn \\[4pt]
& \hspace{1cm} = \frac{3}{2} \, \int_{k_1,k_2,k_3} \, V_3(k_1,k_2,k_3) \, \bigl( \langle \tte^{k_3}, \sgreen(\tte^{p_1})\rangle_\star \,  \bigl\{\CS_{\rm int},\sH\, (\tte^{k_1}\odot_\star \tte^{k_2}\odot_\star \tte^{p_2}) \bigr\}_\star \, \nn\\
& \hspace{7cm} + \langle \tte^{k_3}, \sgreen(\tte^{p_2})\rangle_\star \,\bigl\{\CS_{\rm int},\sH\, (\tte^{p_1}\odot_\star \tte^{k_1}\odot_\star \tte^{k_2}) \bigr\}_\star \bigr)\nn \ .
\end{align}
The first term is given explicitly by
\begin{align}
& \bigl\{\CS_{\rm int},\sH\, (\tte^{k_1}\odot_\star \tte^{k_2}\odot_\star \tte^{p_2}) \bigr\}_\star \nn \\[4pt]
& \hspace{1cm} = \frac{3}{3} \, \int_{q_1,q_2,q_3} \, V_3(q_1,q_2,q_3) \, \bigl( \langle \tte^{q_3}, \sgreen(\tte^{k_1})\rangle_\star \, \tte^{q_1}\odot_\star \tte^{q_2}\odot_\star \tte^{k_2} \odot_\star \tte^{p_2} \, \nn\\
& \hspace{6cm} + \langle \tte^{q_3}, \sgreen(\tte^{k_2})\rangle_\star \, \tte^{k_1}\odot_\star \tte^{q_1}\odot_\star \tte^{q_2} \odot_\star \tte^{p_2} \, \nn\\
& \hspace{7cm} + \langle \tte^{q_3}, \sgreen(\tte^{p_2})\rangle_\star \, \tte^{k_1}\odot_\star \tte^{k_2}\odot_\star \tte^{q_1} \odot_\star \tte^{q_2} \bigr) \ , \nn
\end{align}
and similarly for the second term.

The next step is to apply the operator $\ii\,\hbar\,\BVL\,\sH$ twice in a row, which is equivalent to the braided Wick theorem applied to any product of four fields found under the integral. The result is given by
\begin{align}
&(\ii\,\hbar\,\BVL\,\sH)^2 \, \bigl\{ \CS _{\rm int},\sH\, \bigl\{\CS _{\rm int},\sH\,(\tte^{p_1}\odot_\star \tte^{p_2})\bigr\}_\star \bigr\}_\star \nn \\[4pt]
& \quad = \frac{3}{2} \, \int_{k_1, k_2, k_3} \ \int_{q_1, q_2, q_3} \, V_3(k_1,k_2,k_3) \, V_3(q_1,q_2,q_3) \, \Bigl( \langle \tte^{k_3}, \sgreen(\tte^{p_1})\rangle_\star \, \langle \tte^{q_3}, \sgreen(\tte^{k_1})\rangle_\star\nn \\
&\hspace{3cm} \times \bigl( \langle \tte^{q_1}, \sgreen(\tte^{q_2})\rangle_\star\,\langle \tte^{k_2}, \sgreen(\tte^{p_2})\rangle_\star + \e^{\,\ii\,k_2\cdot\theta\,p_2}\, \langle \tte^{q_1}, \sgreen(\tte^{k_2})\rangle_\star\,\langle \tte^{q_2}, \sgreen(\tte^{p_2})\rangle_\star \nn \\
& \hspace{8cm} + \langle \tte^{q_1}, \sgreen(\tte^{p_2})\rangle_\star\, \langle \tte^{q_2}, \sgreen(\tte^{k_2})\rangle_\star\bigr) + \dots \Bigr) \ ,\nn
\end{align}
where for brevity we display explicitly only the braided Wick expansion of the first product arising.

\begin{figure}[h]%
\centering
\includegraphics[scale=0.7]{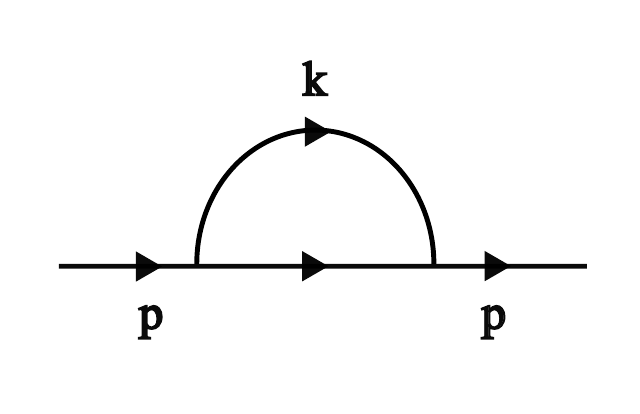}
\vspace{-5mm}
\caption{\small The vacuum polarization contribution to the two-point function at one-loop in braided $\lambda\,\phi_6^3$-theory.}
\label{fig:Phi32}
\end{figure}

Unravelling everything, one gets the second correction term 
\begin{align} \label{DrugaKorekcija}
\begin{split}
\tilde G_2^\star(p_1,p_2)_2^{(1)} &= \frac{(\ii\,\hbar\,\lambda)^2}{12}\, \frac{(2\pi)^6\, \delta(p_1)}{p_1^2-m^2} \ \left[\int_{k}\, \frac{1}{k^2-m^2} \right]^2 \ \frac{(2\pi)^6\, \delta(p_2)}{p_2^2-m^2} \\
& \quad\, +\frac{(\ii\,\hbar\,\lambda)^2}{6}\, \frac{(2\pi)^6\, \delta(p_1+p_2)}{(p_1^2 - m^2)\,(p_2^2 - m^2)} \ \int_k\, \frac{1}{(0-m^2)\,(k^2 - m^2)}\\
& \quad\, +\frac{(\ii\,\hbar\,\lambda)^2}{2}\, \frac{(2\pi)^6\, \delta(p_1+p_2)}{(p_1^2 - m^2)\,(p_2^2 - m^2)} \ \int_k\, \frac{1}{\big(k^2-m^2\big)\,\big((p_1-k)^2 - m^2\big)}  \ .
\end{split}
\end{align}
The first and second terms are the same as the two terms we found in $\tilde G_2^\star(p_1,p_2)^{(1)}_1$. The third term in \smash{$\tilde G_2^\star(p_1,p_2)^{(1)}_2$} comes from a vacuum polarization diagram, see Figure~\ref{fig:Phi32}.

Finally, by adding the two corrections one finds that the one-loop contribution to the two-point function in momentum space is given by
\begin{align} \label{KorekcijaPhi3}
\begin{split}
\tilde G_2^\star(p_1,p_2)^{(1)} &= -\frac{(\hbar\,\lambda)^2}{4} \, \frac{(2\pi)^6\, \delta(p_1)}{p_1^2-m^2} \ \left[\int_{k}\, \frac{1}{k^2-m^2} \right]^2 \ \frac{(2\pi)^6\, \delta(p_2)}{p_2^2-m^2} \\
& \quad \, -\frac{(\hbar\,\lambda)^2}{2} \, \frac{(2\pi)^6\, \delta(p_1+p_2)}{(p_1^2 - m^2)\,(p_2^2 - m^2)} \ \int_k\, \frac{1}{(0-m^2)\,(k^2 - m^2)}\\
& \quad \, -\frac{(\hbar\,\lambda)^2}{2} \, \frac{(2\pi)^6\, \delta(p_1+p_2)}{(p_1^2 - m^2)\,(p_2^2 - m^2)} \ \int_k\, \frac{1}{\big(k^2-m^2\big)\,\big((p_1-k)^2 - m^2\big)} \ .
\end{split}
\end{align}
This is exactly the same result as in the commutative case because braided features that come from different places, such as the vertex and the braided Wick theorem, have cancelled each other out. Additionally, all numerical factors agree with the commutative result. Therefore, we conclude that also in the braided noncommutative $\phi^3$-theory in six dimensions, UV/IR mixing is absent from the two-point function at one-loop. 

It would be interesting to further extend our analysis to calculate the one-loop beta-function for this model and to check whether the braided noncommutative $\phi^3$-theory in six dimensions is still asymptotically free. It would also be interesting to explore the extent to which our conclusions concerning the absence of non-planar diagrams and UV/IR mixing prevail at higher loops and in higher-point correlation functions, in both the $\phi^3$ and $\phi^4$ scalar field theories. We plan to investigate these and other problems in future work.

\appendix

\renewcommand{\theequation}{\Alph{section}.\arabic{equation}}
\setcounter{equation}{0}

\section{Drinfel'd twist deformation formalism}
\label{app:Drinfeld}

In this appendix we briefly introduce the basics of Drinfel'd twist deformation formalism and the notation we use in the main text. More detailed and thorough treatments can be found in~\cite{MajidBook, SLN}.

In the Drinfel'd twist formalism, a deformation is introduced via a twist element ${\cal F} \in U\frv \otimes U\frv$, where $U\frv$ is the universal enveloping algebra of the Lie
algebra of vector fields $\frv:=\Gamma(TM)$ on a manifold $M$.  We use the notation ${\cal F} = \rm{f}^\alpha\otimes \rm{f}_\alpha$ and ${\cal F}^{-1} =  \bar{\rm{f}}^\alpha\otimes 
\bar{\rm{f}}_\alpha$. 

The invertible $\RR$-matrix $\RR\in
U\frv\otimes U\frv$ encodes the braiding and it is induced by the twist as
\begin{align}\nn
\RR=\mathcal{F}_{21}\, {\cal F}^{-1}=:\sfR^\alpha\otimes\sfR_\alpha \ ,
\end{align}
where $\mathcal{F}_{21}=\tau(\mathcal{F})=\mathrm{f}_\alpha\otimes\mathrm{f}^\alpha$ is the twist with its legs swapped. It is easy to see that the $\RR$-matrix is triangular, that is
\begin{equation}
  \RR_{21} = \RR^{-1} = \sfR_\alpha\otimes\sfR^\alpha \ .\label{Inv_R}
\end{equation}

The simplest class of twists are the abelian twists which are of the form
\begin{equation}
{\cal F} =  \exp\big(-\tfrac{\ii}{2}\,\theta^{ab}\,X_a \otimes 
X_b\big) \  ,\label{Abelian_Twist} 
\end{equation}
where $(\theta^{ab})$ is a constant antisymmetric matrix and $\{X_{a}\}$ is a set of commuting vector fields. For these twists $\CF_{21}=\CF^{-1}$ and  $\RR=\CF^{-2}$. The standard Moyal--Weyl twist and also the angular twist of~\cite{DimitrijevicCiric:2018blz} are examples of abelian twists. 

The twist (\ref{Abelian_Twist}) deforms the pointwise product $f\cdot g$ on the algebra of functions $C^\infty(M)$ to the noncommutative star-product
\begin{align}
\begin{split}
f\star g &=\> \bar{\rm{f}}^\alpha(f)\cdot \bar{\rm{f}}_\alpha(g) \label{fstarg} = \sfR_\alpha( g)\star \sfR^\alpha( f) \\[4pt]
&=\> f\cdot g + \tfrac{\ii}{2}\,\theta^{ab}\,X_a( f) \cdot X_b (g) + O(\theta^2) \ .  
\end{split}
\end{align}

Extending the action of $U\frv$ to tensor fields using the Lie derivative, the exterior algebra of differential forms $\Omega^\bullet(M)$ is deformed in a similar way with
\begin{align}
& \omega_1\wedge_\star\omega_2 = \bar{\rm{f}}^\alpha(\omega_1)\wedge \bar{\rm{f}}_\alpha(\omega_2) = (-1)^{|\omega_1|\,|\omega_2|} \, \sfR_\alpha(\omega_2)\wedge_\star \sfR^\alpha(\omega_1) \ ,\label{StarWedge}
\end{align}
and
\begin{align}
& \d(\omega_1\wedge_\star\omega_2) = \d\omega_1\wedge_\star\omega_2 + (-1)^{|\omega_1|}\,
\omega_1\wedge_\star\d \omega_2  \ .\nn
\end{align}
By $|\omega|$ we denote the degree of a homogeneous form $\omega$. The exterior derivative $\d$ does not change under the deformation.

We define braided field theories using the twist formalism. A classical field theory on $\FR^{1,d-1}$ is generically Poincar\'e invariant, so its $L_\infty$-algebra consists of modules and equivariant brackets for the universal  enveloping algebra $U\mathfrak{iso}(1,d-1)\subset U\frv$ of the Poincar\'e algebra $\mathfrak{iso}(1,d-1)$. Hence we have to restrict to twists $\CF\in U\iso(1,d-1)\otimes U\iso(1,d-1)$. We can further restrict to abelian twists $\CF\in U\iso(d-1)\otimes U\iso(d-1)$ constructed from the spatial isometries of $\FR^{d-1}\subset\FR^{1,d-1}$, as this simplifies some of the analysis in the quantum field theory, such as the treatment of time-ordering, as well as avoiding potential issues with unitarity. 

For definiteness, and for the sake of illustration, in the main text we choose the Moyal--Weyl twist
\begin{align}\label{eq:MWtwist1}
\CF =\exp\big(-\tfrac{\mathrm{i}}2\,\theta^{ij}\,\partial_i\otimes\partial_j\big) 
  \ ,
\end{align}
where $(\theta^{ij})$ is a $(d-1){\times}(d-1)$ antisymmetric real-valued
matrix, and $\partial_i=\frac\partial{\partial x^i}\in\Gamma(T\FR^{d-1})$ for $i=1,\dots,d-1$ are vector fields generating spatial translations in \smash{$\FR^{1,d-1}$}.

\paragraph{Acknowledgments.}
We thank the organisers of the Corfu Summer Institute
2022 for the stimulating meeting and the opportunity to present the
preliminary results of our work. We also thank Nikola Konjik and Biljana Nikoli\'c for helpful discussions.
The work of {\sc Dj. B., M.D.C.} and  {\sc V.R.} is supported by Project
451-03-47/2023-01/ 200162 of the Serbian Ministry of Education, Science and
Technological Development. The work of {\sc R.J.S.} was supported in part by
the Consolidated Grant ST/P000363/1 
from the UK Science and Technology Facilities Council.


\begin{thebibliography}{99}

\bibitem{Minwalla:1999px}
S.~Minwalla, M.~Van Raamsdonk and N.~Seiberg,
{\it Noncommutative perturbative dynamics},
JHEP \textbf{02} (2000) 020, [arXiv:hep-th/9912072].

\bibitem{U1Reviewa}
R. J. Szabo, {\it Quantum field theory on noncommutative spaces}, Phys.\ Rept. {\bf 378} (2003) 207--299, [arXiv:hep-th/0109162].

\bibitem{Blaschke}
D. Blaschke, {\it Aspects of perturbative quantum field theory on noncommutative spaces},  Proc. Sci. {\bf 263} (2016) 104, [arXiv:1601.03109].

\bibitem{Bal}
A.~P.~Balachandran, A.~Pinzul and B.~A.~Qureshi,
{\it UV/IR mixing in noncommutative plane},
Phys. Lett. B \textbf{634} (2006) 434--436,
[arXiv:hep-th/0508151].

\bibitem{Bu}
J.~G.~Bu, H.~C.~Kim, Y.~Lee, C.~H.~Vac and J.~H.~Yee,
{\it Noncommutative field theory from twisted Fock space},
Phys. Rev. D \textbf{73} (2006) 125001, [arXiv:hep-th/0603251].

\bibitem{Wess}
G.~Fiore and J.~Wess,
{\it On full twisted Poincar\'e symmetry and QFT on Moyal--Weyl spaces},
Phys. Rev. D \textbf{75} (2007) 105022, [arXiv:hep-th/0701078].

\bibitem{Aschieri:2007sq}
P.~Aschieri, F.~Lizzi and P.~Vitale,
{\it Twisting all the way: From classical mechanics to quantum fields}, Phys. Rev. D \textbf{77} (2008) 025037, [arXiv:0708.3002].

\bibitem{Oeckl}
R. Oeckl, {\it Untwisting noncommutative $\mathbb{R}^d$ and the equivalence of quantum field theories}, Nucl.\ Phys.\ B {\bf 581} (2000) 559--574 [arXiv:hep-th/0003018].

\bibitem{BraidedLinf}
M.~Dimitrijevi\'c~\'Ciri\'c, G.~Giotopoulos, V.~Radovanovi\'c and R.~J.~Szabo,
{\it Braided $L_{\infty}$-algebras, braided field theory and noncommutative gravity}, Lett.\ Math.\ Phys.\ {\bf 111} (2021) 148, [arXiv:2103.08939].

\bibitem{BVChristian} 
B.~Jur\v{c}o, L.~Raspollini, C.~S\"amann and M.~Wolf,
{\it $L_\infty$-algebras of classical field theories and the Batalin--Vilkovisky formalism},
Fortsch.\ Phys.\  {\bf 67} (2019) 1900025,
[arXiv:1809.09899].

\bibitem{SzaboAlex}
H. Nguyen, A.  Schenkel and R. J.  Szabo,  {\it Batalin--Vilkovisky quantization of fuzzy field theories}, Lett. Math. Phys. {\bf 111} (2021) 149, [arXiv:2107.02532].

\bibitem{UsQED}
M.~Dimitrijevi\'c~\'Ciri\'c, N. Konjik, V.~Radovanovi\'c and R.~J.~Szabo, {\it Braided quantum electrodynamics}, arXiv:2302.10713.

\bibitem{HohmZwiebach}
O.~Hohm and B.~Zwiebach, {\it $L_{\infty}$-algebras and field theory},
Fortsch. Phys. \textbf{65} (2017) 1700014, [arXiv:1701.08824].

\bibitem{Gomis94}
J. Gomis, J. Paris and S. Samuel, {\it Antibracket, antifields
and gauge theory quantization}, Phys.\ Rept. {\bf 259} (1995) 1--145, [arXiv:hep-th/9412228].

\bibitem{Doubek:2017naz}
M.~Doubek, B.~Jur\v{c}o and J.~Pulmann,
{\it Quantum $L_\infty$-algebras and the homological perturbation lemma},
Commun. Math. Phys. \textbf{367} (2019) 215--240, [arXiv:1712.02696].

\bibitem{Giotopoulos:2021ieg}
G.~Giotopoulos and R.~J.~Szabo,
{\it Braided symmetries in noncommutative field theory},
J. Phys. A \textbf{55} (2022) 353001, [arXiv:2112.00541].

\bibitem{MajidBook}
S.~Majid, {\it Foundations of Quantum Group Theory}, Cambridge University Press (1995).

\bibitem{SLN}
P.~Aschieri, M.~Dimitrijevi\' c, P.~Kulish, F.~Lizzi and J.~Wess, {\it Noncommutative spacetimes: Symmetries in noncommutative geometry and field theory},
Lect.\ Notes Phys. \textbf{774} (2009) 1--199.

\bibitem{DimitrijevicCiric:2018blz}
M.~Dimitrijevi\'c \'Ciri\'c, N.~Konjik, M.~A.~Kurkov, F.~Lizzi and P.~Vitale, {\it Noncommutative field theory from angular twist},
Phys. Rev. D \textbf{98} (2018) 085011, [arXiv:1806.06678].

\end{thebibliography}
\end{document}